%% file: 0A_MAIN.tex
\documentclass[lettersize,journal]{IEEEtran}
\usepackage{algorithmic}
\usepackage{array}
\usepackage{textcomp}
\usepackage{stfloats}
\usepackage{url}
\usepackage{verbatim}
\usepackage{graphicx}
\usepackage{cite}
\usepackage{tikz}
\usepackage{pgfplots}
\pgfplotsset{width=10cm,compat=1.9}
\usepackage{amsmath,amssymb,amsfonts}
\usepackage{url}

\usepackage{subcaption}
\usepackage[export]{adjustbox}
\usepackage{capt-of}
\usepackage{caption}
\usepackage{booktabs}
\usepackage{float}
\usepackage{multirow}
\usepackage{enumitem}
\usepackage[export]{adjustbox}
\usepackage[linesnumbered,ruled,vlined]{algorithm2e}

\begin{document}

\bstctlcite{IEEEexample:BSTcontrol}

\title{Network Sovereignty: A Novel Metric and its Application on Network Design}

\author{Shakthivelu Janardhanan,~\IEEEmembership{Student Member,~IEEE,}
Maria Samonaki,~\IEEEmembership{Student Member,~IEEE,}
Poul Einar Heegaard,~\IEEEmembership{Senior Member,~IEEE,}
Wolfgang Kellerer,~\IEEEmembership{Senior Member,~IEEE,}
Carmen Mas-Machuca,~\IEEEmembership{Senior Member,~IEEE.}


\thanks{This work has been funded by the Bavarian Ministry of Economic Affairs, Regional Development, and Energy under the project "6G Future Lab Bavaria", by the Federal Ministry of Education and Research of Germany in the program of “Souver\"an. Digital. Vernetzt.”. Joint project 6G-life, project identification number: 16KISK002 and the German Research Foundation
(DFG) under grant numbers MA 6529/4-1 and KE 1863/10-1}
}
\markboth{Journal of \LaTeX\ Class Files,~Vol.~14, No.~8, August~2021}%
{Shell \MakeLowercase{\textit{et al.}}: A Sample Article Using IEEEtran.cls for IEEE Journals}


\maketitle

\input{0B_Abstract}
\input{1_Introduction}
\input{2_Related_work}
\input{3A_Challenges}
\input{4_Metric}
\input{5_Naga}
\input{5b_evalworkflow}
\input{6_Results}

\input{7_Conclusion}

\bibliographystyle{IEEEtran}
\bibliography{IEEEabrv, 2_Bibliography}


\section{Biography Section}


\vspace{-30pt} 
\begin{IEEEbiography}[{\includegraphics[width=1in,height=1.25in,clip,keepaspectratio]{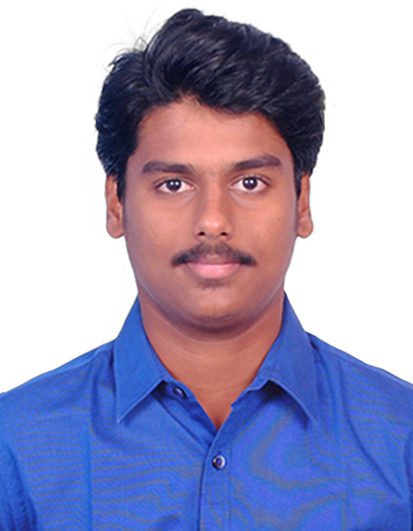}}]{Shakthivelu Janardhanan}
received his Bachelor's degree in Electronics and Communication Engineering from SSN College of Engineering, India, in 2019. He received his Master of Science degree in Communication Engineering from the Technical University of Munich in November 2021. In December 2021, he joined the Chair of Communication Networks at TUM as a doctoral candidate. His research interests include network reliability, network sovereignty, and modeling and analysis of networks.
\end{IEEEbiography}
\begin{IEEEbiography}[{\includegraphics[width=1in,height=1.25in,clip,keepaspectratio]{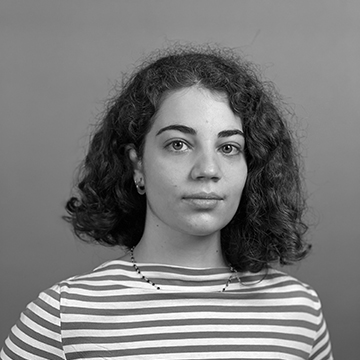}}]{Maria Samonaki}
received her Diploma degree (Integrated Master's) in Electrical and Computer Engineering from the Technical University of Crete, Chania, Greece, in March 2022. In April 2022, she joined the Chair of Communication Networks at TUM as a doctoral candidate. Her research interests lie in reliability in multi-domain networks and optimal reliable network planning. 
\end{IEEEbiography}
\begin{IEEEbiography}[{\includegraphics[width=1in,height=1.25in, trim = {3cm 0 0 0}, clip]{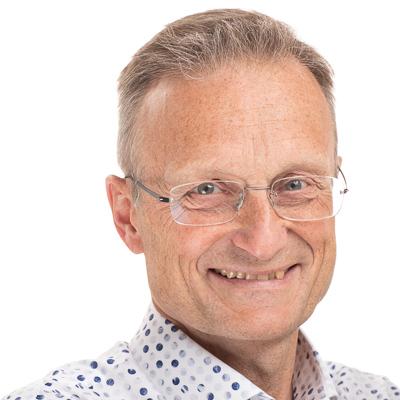}}]{Poul Einar Heegaard} is Full Professor at the Department of Information Security and Communication Technology, Norwegian University of Science and Technology (NTNU), where he also has acted both as head of department and head of the research group in Networking. He was previously Senior Scientist with SINTEF Digital (1989-1999) and then Telenor R\&I (1999-2009). He is a senior member of the IEEE.
\end{IEEEbiography} 
\begin{IEEEbiography}[{\includegraphics[width=1in,height=1.25in, trim = {0 5cm 0 2cm}, clip]{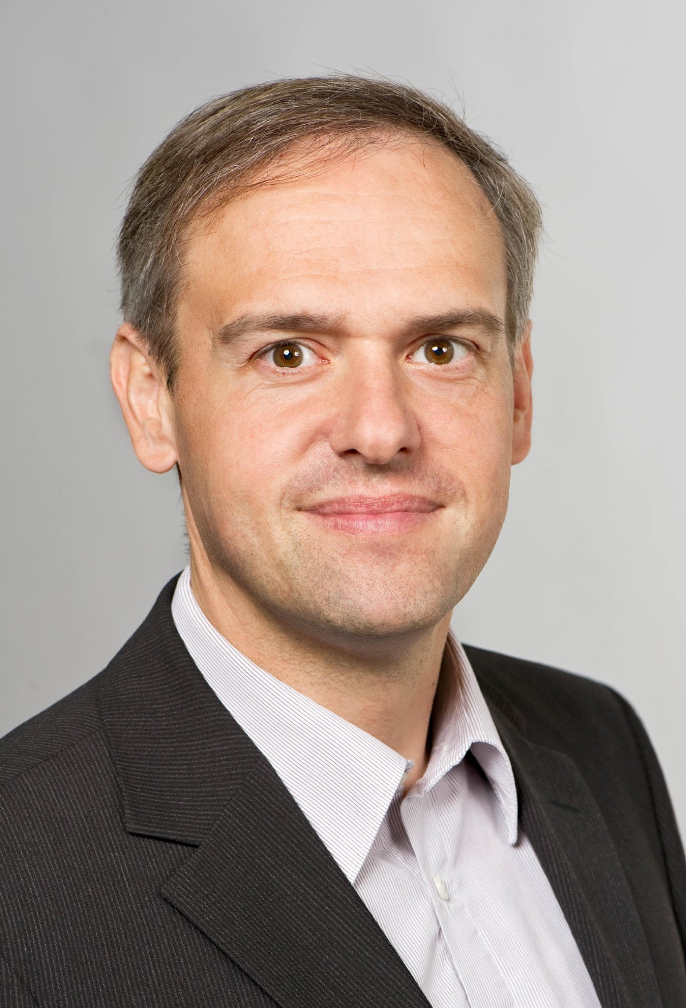}}]{Wolfgang Kellerer}
is a Full Professor at the Technical University of Munich (TUM), heading the Chair of Communication Networks. Before, he was for over ten years with NTT DOCOMO’s European Research Laboratories. He currently serves as an associate editor for IEEE Transactions on Network and Service Management and as an area editor for Network Virtualization for IEEE Communications Surveys and Tutorials.
\end{IEEEbiography} 
\begin{IEEEbiography}[{\includegraphics[width=1in,height=1.1in,trim = {20cm 4cm 20cm 0}, clip]{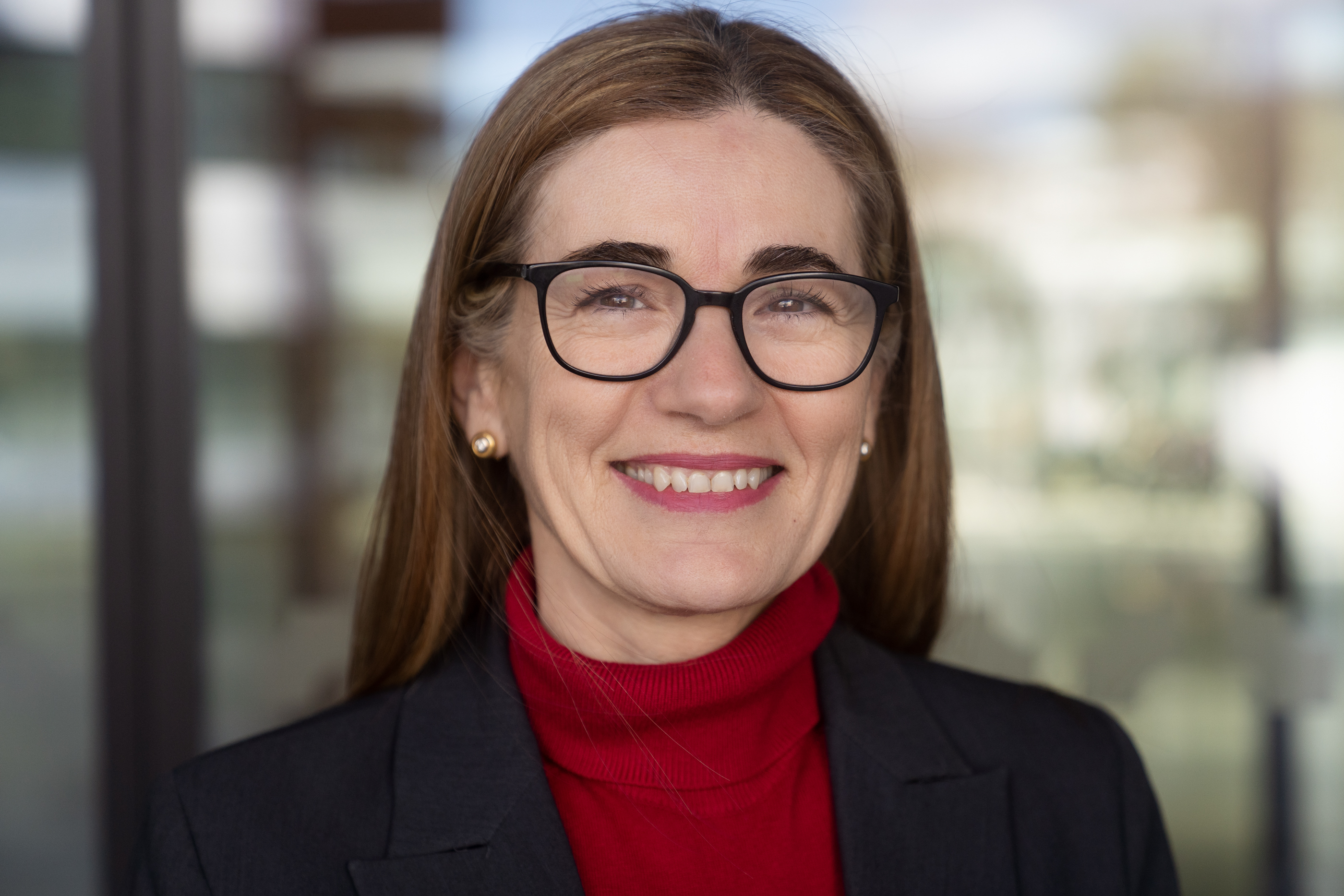}}]{Carmen Mas-Machuca}
is a Full professor with the Chair of Communication Networks at the University Bundeswehr München (UniBW), Germany. Her research interests include techno-economic studies, network planning and resilience, optimization problems, and optical networks. She is a guest editor for IEEE Transactions on Network and Service Management, OSA Journal on Optical Communications and Networking, and IEEE Communications Magazine and is also active in international conferences as chair, TPC co-chair, and TPC member.
\end{IEEEbiography}

\vspace{11pt}

\vfill

\end{document}

%% file: 0B_Abstract.tex
\begin{abstract}

Most network planning problems in literature consider metrics such as cost, availability, and other technology-aware attributes. However, network operators now face new challenges in designing their networks to minimize their dependencies on manufacturers. 
A low dependency is associated with higher network robustness in case one or more manufacturers fail due to erroneous component design, geopolitical banning of manufacturers, or other reasons discussed in this work. 
Our work discusses network sovereignty, i.e., the ability to operate a network while minimizing the dependencies on a particular manufacturer to minimize the impact of simultaneous manufacturer failure(s). 
Network sovereignty is considered by solving the manufacturer assignment problem in the network such that robustness is maximized.
The three main contributions of this work are (i)~the discussion of network sovereignty as a special attribute of dependability, (ii)~the introduction of a novel metric- the Path Set Diversity (PSD) score to measure network sovereignty, and (iii)~the introduction of \textit{Naga}, an Integer Linear Program (ILP) formulation to maximize network sovereignty using the PSD score. We compare \textit{Naga}'s performance with centrality metrics-based heuristics and an availability-based optimization. Our work aims to be the foundation to guide network operators in increasing their network sovereignty.

\end{abstract}
\begin{IEEEkeywords}
network sovereignty, dependability, path set diversity, measurement metric, manufacturer failure
\end{IEEEkeywords}

%% file: 1_Introduction.tex
\section{Introduction}
\label{chap:intro}

A healthy and fair market usually has multiple manufacturers supplying comparable products. 
Network operators prefer to buy most components from one or two manufacturers since purchase and deployment in bulk is cheaper, while interoperability is simpler.
However, purchasing components from one or two manufacturers creates a strong dependency on those manufacturers, similar to the vendor lock-in problem. In an unforeseen circumstance, if the manufacturer is unavailable or banned, all of that manufacturer's components will be unavailable, affecting both the operator and the customers. 

Several incidents have occurred where all components from one manufacturer have been affected simultaneously. For example, the Samsung Galaxy Note 7 unit explosion~\cite{samsung} forced Samsung to recall 2.5 million units in September 2016. The analysis revealed a manufacturing defect in the batteries. Moreover, incidents could also be software-related. For example, 4000 government websites were temporarily out of service in Canada in December 2021 due to a vulnerability reported in Apache Log4j- a Java-based logging utility~\cite{canada}. The Equifax data leak in mid-2017~\cite{equifax}, caused by some servers using an outdated Apache Struts library, compromised millions of users' data. 
A significant downtime was caused in all these examples, leading to high monetary losses.

Such simultaneous failures can not be effectively described using traditional dependability attributes~\cite{Avizienis:2004} such as availability, reliability, safety, integrity, and maintainability. This warrants the need for another attribute that addresses the effect of manufacturer dependency and trustworthiness. 
Consequentially, technology sovereignty was proposed~\cite{edler2020technology, weber2018sovereignty} and defined as the ability of an organization to provide a technology by developing it or outsourcing it efficiently without causing a dependency on a particular supplier. However, developing all technologies entirely indigenously is impractical for any organization. Therefore, technology sovereignty's main focus is on removing dependencies on external suppliers.

Our work focuses on \textit{network sovereignty}, which is qualitatively defined as an organization's ability to operate a network while minimizing the dependencies on a particular manufacturer. 
An `organization' refers to any institution that operates independently, whereas the `manufacturer' (or a vendor) could be a hardware manufacturer, software company, service provider, raw material supplier, or any part of the supply chain leading to the development of the network components. 

\begin{center}
\begin{figure}[t]
\begin{subfigure}{0.34\linewidth}
\includegraphics[width=1\linewidth, center]{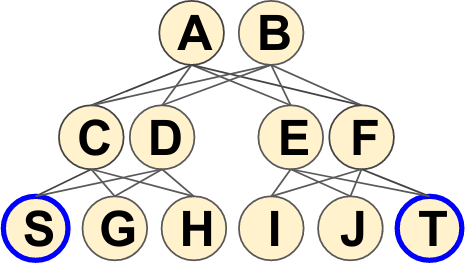} 
\caption{1 manufacturer.}
\label{fig:thediff_A}
\end{subfigure}
\begin{subfigure}{0.6\linewidth}
\includegraphics[width=1\linewidth, center]{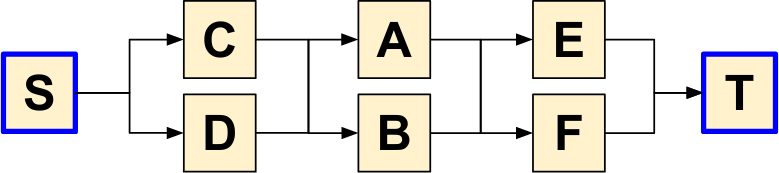} 
\caption{Reliability Block Diagram for (a).}
\label{fig:thediff_B}
\end{subfigure}
\begin{subfigure}{0.34\linewidth}
\includegraphics[width=1\linewidth, center]{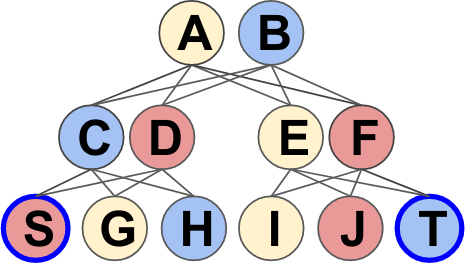} 
\caption{Example manufacturer assignment for 3 manufacturers.}
\label{fig:thediff_C}
\end{subfigure}
\begin{subfigure}{0.6\linewidth}
\includegraphics[width=1\linewidth, center]{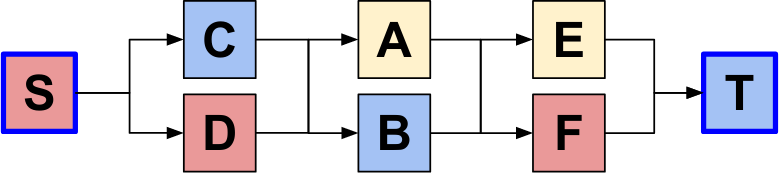} 
\caption{Reliability Block Diagram for (c).}
\label{fig:thediff_D}
\end{subfigure}
\caption{Availability and Sovereignty comparison of a network with 1 and 3 different manufacturers.}
\label{fig:thediff}
\end{figure}
\end{center} 

Network sovereignty is often confused with network dependability attributes like availability~\cite{Avizienis:2004}. To understand the difference, let us consider an example network of 12 interconnected nodes, as shown in Fig.~\ref{fig:thediff_A}, with all nodes to be from the same yellow manufacturer with a node availability of 0.999 and a flow between the nodes $S$ and $T$. 
The Reliability Block Diagram (RBD) of this flow is shown in Fig.~\ref{fig:thediff_B}, and by analyzing the parallel and serial connections, the flow availability is 0.998. However, none of the nodes are available if this single manufacturer is unavailable. This implies that the flow can not be routed; hence, the network is not sovereign as it totally depends on this single manufacturer.
Let us now consider having three different manufacturers in the same network- depicted as red, yellow, and blue manufacturers in Fig.~\ref{fig:thediff_C}, having node availabilities of 0.9985, 0.999, and 0.9995, respectively. In this case, the flow availability is still 0.998. However, sovereignty has improved because even when all the nodes from the yellow manufacturer fail, there are still multiple paths between $S$ and $T$. 
Therefore, introducing more manufacturers into the network has successfully removed the structural dependency on the yellow manufacturer. 
Additionally, this example shows how traditional dependability attributes like availability cannot explain and prevent multiple simultaneous failures from a single manufacturer.

The first major challenge in network sovereignty is the absence of a metric. The absence of metrics makes it impossible to measure and compare the sovereignty of different manufacturer assignments or networks. Therefore, as a first step toward measuring network sovereignty, we present this paper with the following contributions.
\begin{enumerate}
\item Discuss network sovereignty as a possible special attribute of dependability (Section~\ref{chap:intro}).
\item Identify the major challenges in building a sovereign network (Section~\ref{chap:challenges}).
\item Introduce `Path Set Diversity (PSD) score'- the first novel metric to measure network sovereignty (Section~\ref{chap:metric}).
\item Introduce the Manufacturer Assignment for Sovereignty (MAS) problem and then solve it using \textit{Naga}, our ILP formulation based on the PSD score (Section~\ref{chap:Naga}).
\item Evaluate \textit{Naga} and compare its performance with (i)~centrality metrics-based heuristics like Nodal degree, Betweenness centrality, and Closeness centrality, and (ii)~\textit{Zohra}, the state of the art on Manufacturer Assignment for Availability (Sections~\ref{chap:evalworkflow} and~\ref{chap:results}).
\end{enumerate}

%% file: 2_Related_work.tex
\section{Related Work}
\label{chap:bg}

Sovereignty, defined in 1577~\cite{bodin}, has evolved over time and continues to be relevant. In the late 20th century, sovereignty intertwined with the digital realm~\cite{grant}. This created new ideas and interpretations of data, digital, and technological sovereignty. 
Technology sovereignty~\cite{edler2020technology,weber2018sovereignty} and data sovereignty~\cite{review_data_sovereignty} are well-defined in literature.
Most previous works are on international governance and policy-making, primarily focusing on providing a secure internet by ensuring data privacy~\cite{sov2, sov1, sov3, sov4} and internet filtering~\cite{asia2}. They also focus on the importance of indigenous sovereignty~\cite{walter2021indigenous, carroll2019indigenous}, data localization~\cite{hill2014growth}, and the interpretations of trade policies~\cite{gao2021data}.
Several countries are focussing on data sovereignty. For example, the authors in~\cite{niceone_cyberspace, niceone_cyberspace2} emphasize the approaches to sovereignty from the BRICS nations (Brazil, Russia, India, China, and South Africa) and their defiance against the United States of America's (USA) internet domination. 
In~\cite{chander2014data_nationalism}, the authors summarize the data nationalism policies of 17 countries and regions.
For example, Russia and China store citizens' data inside the nation only, while others like the EU and Brazil have been considering this.

Authors in~\cite{litsurveysovereignty} study the different terms- data sovereignty, digital sovereignty, and technological sovereignty. They conclude that (i)~data sovereignty focuses on Information Systems activities by protecting data on organizational and individual levels, (ii)~digital sovereignty focuses on interoperability, and (iii)~technology sovereignty is the umbrella term for international rules and regulations. However, they do not discuss network sovereignty.
Network sovereignty requires guaranteeing (i)~protection of data assets, (ii)~interoperability among manufacturers, (iii)~reliable network control and operation, (iv)~favorable programmable hardware components, and (v)~flexible, modular software components. 


Two terms worthy of discussion are network redundancy and network diversity~\cite{red, div, redvsdiv}. Network redundancy refers to the duplication of network elements. This guarantees multiple paths between a source and a destination. For example, most campus networks have active WiFi and Ethernet connectivity. This gives the user two different means of communicating with the campus network and the internet. Such a redundancy improves reliability. However, if both connections are from the same vendor, there is no improvement in network sovereignty. 

On the other hand, network diversity is an improvement in redundancy. Network diversity refers to the duplication of different levels of the network infrastructure to provide multiple technology/vendor-independent paths between the source and destination. Not to be confused with geographically diverse routing, network diversity refers to having multiple independent network providers to avoid a dependency on a single network provider. Such a concept is already being advertised by internet service solutions companies like Astound-Grande~\cite{astound}.
However, network sovereignty aims to provide the advantages of network diversity without duplicating network infrastructure, thereby saving costs and resources.


Looking into implementation-oriented works on multi-vendor networks, \cite{bauschert_vendor_selection, AugeOFC, gabilondo20215g, multi_vendor} provide great insight. 
In~\cite{bauschert_vendor_selection}, the authors examine the impact of selecting manufacturers on the total cost of ownership (TCO) of data center networks (DCN). 
In \cite{AugeOFC}, the focus is on optical physical layer models to design open-line systems, enabling vendor interoperability. 
In \cite{gabilondo20215g}, an end-to-end multi-vendor 5G Standalone mobile network setup is assessed for interoperability. 
In \cite{multi_vendor}, the authors examine the interoperability challenges in multi-layer multi-vendor carrier-grade networks.
Though these state of the art works on multi-vendor networks do not discuss sovereignty or reliability, they lay an excellent foundation for considering the possibility and strategy behind using multiple vendors in a network.

Our previous work~\cite{mine4} discusses a solution to the Joint Routing and Manufacturer Assignment (JRMA) problem, which aims to find the best manufacturer assignment and routes for the various flows in the network to maximize network availability. The proposed solution was a non-linear integer program called \textit{Zohra}. However, this work strictly adheres to network availability and does not discuss sovereignty. However,~\cite{mine4} is a good reference for the manufacturer assignment problem, and it will be used in the performance evaluation of this work for comparison.

Only a few works focus on `network sovereignty.' 
Our previous works~\cite{mine1, mine2} review different DCN topologies of varying sizes and characterize the usage of multiple manufacturers in building a DCN. In~\cite{mine1, mine2}, we discuss the manufacturer assignment problem to improve network sovereignty and provide guidelines to operators to build a sovereign DCN. To the best of our knowledge, these two works stand as the lone attempt to provide insight into manufacturer assignment strategies for improving network sovereignty from a technical perspective as opposed to the policy papers in the fields of economy and politics.
However, the state of the art can still not quantify network sovereignty. They are only able to provide general guidelines to network operators. 
No work so far has provided a means to measure, compare, or mathematically evaluate a network's sovereignty. For the first time, our work aims to break this barrier and provide a new metric to measure network sovereignty. Additionally, we provide a method to optimize network sovereignty. The differences between the previous works and this work are mentioned in Table~\ref{tab:sota}.

\begin{table}[]
\centering
\caption{Comparison with the State of the Art on the manufacturer assignment problem.}
\label{tab:sota}
\begin{tabular}{|p{2cm}|p{1.6cm}|p{1.6cm}|p{1.6cm}|}
\hline
\textbf{Comparing criteria} &
  \textbf{Work in~\cite{mine4}} &
  \textbf{Work in~\cite{mine1},~\cite{mine2}} &
  \textbf{This work} \\ \hline
Manufacturer Assignment Problem & For availability & For sovereignty                   & For sovereignty          \\ \hline
Metric(s) used                     & Availability & Max-flow and connectivity             & Path Set Diversity score \\ \hline
Results based on                & Non-linear optimization & Trial and error            & Linear optimization      \\ \hline
Focused network                & Any network   & Data center                         & Any network              \\ \hline
How to improve network sovereignty? &
  Not addressed &
  Guidelines based on patterns in the results &

  ILP (\textit{Naga}) to optimize for network sovereignty \\ \hline
\end{tabular}
\end{table}

%% file: 3A_Challenges.tex
\section{Challenges in Network Sovereignty}
\label{chap:challenges}
This section identifies the different challenges in establishing a sovereign network. 
\subsection{Lack of metrics to measure network sovereignty}
\label{sec:metriclack}
We have several parameters to compare the operational efficiency of different networks, such as throughput for performance comparison, round trip delay for latency comparison, availability for dependability comparison, etc. However, there are no metrics for sovereignty. It may be possible to say that one network is more sovereign than the other, as seen in the example in Fig.~\ref{fig:thediff}. However, this is not quantitative. It does not allow the comparison of different manufacturer assignments with the same number of manufacturers in the same network. Even in our previous works~\cite{mine1, mine2}, we only show that one DCN arrangement is more sovereign than the other based on the max-flow observed. This worked for a symmetric topology like the DCN, but it will not work for an asymmetric random topology like a core or wireless network. Hence, we need a metric to measure network sovereignty. This metric must be (i)~well-defined and convenient to calculate, (ii)~scalable to large topologies, and (iii)~consistent across different types of network technologies. This metric is this paper's main focus. 

\subsection{Geopolitical influence}
\label{sec:geopolitical}
Geopolitical influence is another primary challenge. 
For example, China has blocked several services like Google search engine, Gmail, YouTube, Wikipedia, Instagram, WhatsApp, and LinkedIn~\cite{china_website1}. 
On the other hand, in 2020, the U.S. Federal Communications Commission (FCC)~\cite{huawei_US_1} banned Huawei equipment from their networks. 
Following suit, many other countries like the United Kingdom~\cite{huawei_uk_1} and Spain~\cite{spain_china_1} have also tried to reduce Chinese equipment in their networks.
If a network is built with components from a vendor that gets banned by the government, the network operator has to replace them all at their own cost. The operator would be in serious trouble if most components were from the banned vendor.

\subsection{Manufacturer availability}
\label{sec:share}

The reliability of a manufacturer in delivering components is crucial, especially in industries like semiconductors, where integrated chips (ICs) significantly influence hardware cost and performance. Recent trade wars have greatly impacted semiconductor import-export~\cite{uschina_chip1}. For instance, the Taiwan Semiconductor Manufacturing Company (TSMC) has halted IC production for Biren Technology, a Chinese semiconductor company~\cite{uschina_chip2_bloomberg}. Consequently, Biren must reconsider product designs originally intended for TSMC's chips. No company fully develops and processes its raw materials. Therefore, the inter-dependency between different manufacturers raises a question of supplier reliability, leading to 2 different issues:
\subsubsection{Granularity of subcomponents considered} Consider the example of a networking switch that has different subcomponents like Central Processing Unit (CPU), Application-Specific Integrated Circuits (ASIC), etc. Works like \cite{mine3, main_all_values} highlight the importance of such smaller subcomponents from an availability perspective. From a sovereignty perspective, the CPU can have different integrated chips. The integrated chips come from semiconductor foundries. Therefore, the extent to which sovereignty must be considered is still unclear.
\subsubsection{Subcomponent sharing between manufacturers} If a monopoly market dominates an industry, several manufacturers can be expected to use subcomponents from the same monopoly player. For example, TSMC~\cite{tsmc_clients_1, tsmc_clients_2} supplies chips to companies like Apple, NVIDIA, MediaTek, Advanced Micro Devices, Inc. (AMD), Qualcomm, Intel, etc., causing a dependency on TSMC.


\subsection{Multi-domain sovereignty}

\begin{figure*}[htbp]
\centering
\includegraphics[width=0.7\linewidth]{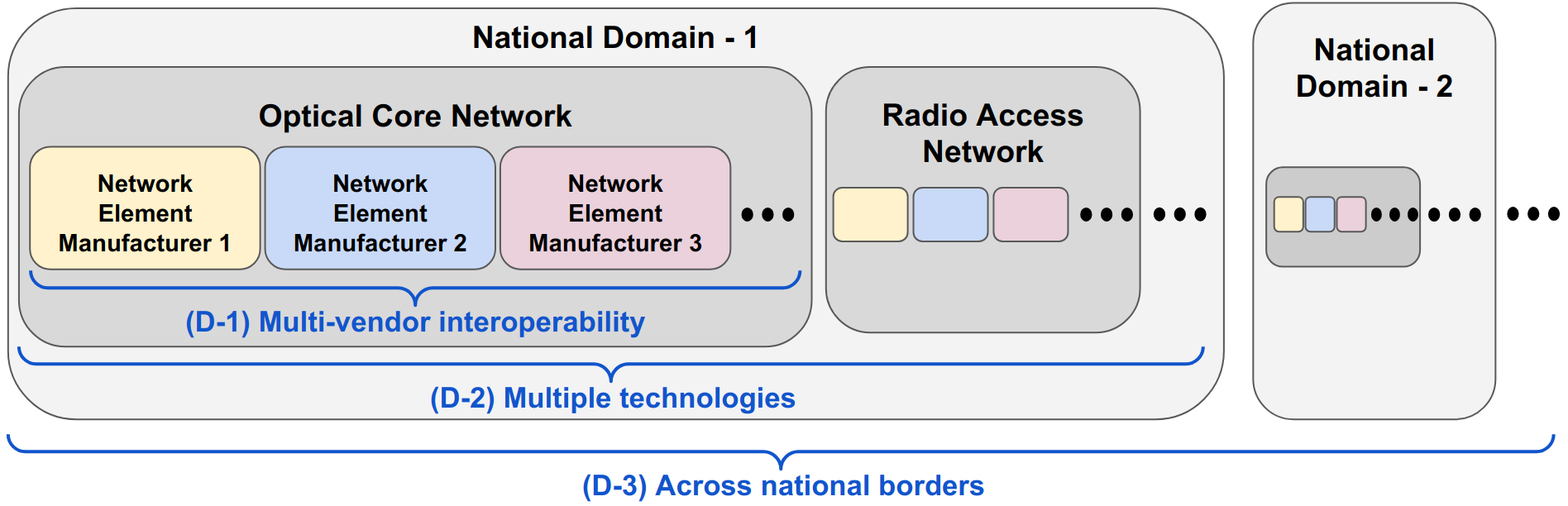}
\caption{Multi-domain reliability and sovereignty perspectives.}
\label{fig:relation}
\end{figure*}

Operators avoid sharing domain-specific information in multi-domain networks for scalability and confidentiality reasons. Hesitation to disclose details such as topology, connectivity, mobility, security, and service availability aims to prevent competitors from assessing business strategies and attackers from exploiting vulnerabilities. The different perspectives on multi-domain sovereignty are depicted in Fig.~\ref{fig:relation} and discussed in the following sections.

\subsubsection{Multi-vendor interoperability} 
\label{sec:disaggregated_network}
Networks with devices from different vendors can be challenging to manage despite all the standardization efforts. Consider the example of an optical core network as seen in Fig.~\ref{fig:relation}.
With the continuous expansion of optical networks, there is always room for upgrading the existing network elements (NEs) with novel functionalities. 
However, if these NEs are from different vendors, interoperability issues such as compatibility between NEs in the data plane and messages between NEs specific to the control plane can be observed~\cite{multi_vendor}. Multi-vendor interoperability is complex in control plane actions because each vendor has solutions complying with the standards but extends to offer their own solutions for network management, communication protocols, and data model definition. This prevents network automation, remote configuration, and dynamic optical service/slice provisioning capabilities. To avoid such issues, service providers are now working on defining open data models like OpenConfig and OpenROADM to move towards a vendor-neutral disaggregated network~\cite{vickybro1}.

\subsubsection{End-to-End (E2E) communication between multiple technologies}
Multi-domain networks can consist of domains managed by different technologies, as seen in Fig.~\ref{fig:relation}. Hence, E2E communication in such heterogeneous networks is a topic of interest. 
For example, the works in~\cite{Bur08, Matos15} focus on provisioning E2E Quality of Service (QoS) in the Internet since the access networks may use different technologies, such as x-Digital Subscriber Line (xDSL), Universal Mobile Telecommunications Service (UMTS), Local Area Network (LAN), WiFi, or Satellite and be connected by many Internet Protocol (IP)-based transit domains. 
In~\cite{Durresi2010}, the authors propose a hierarchical architecture that enables policy-based interconnection, mobility, and other services among domains while discussing heterogeneous multi-domain scalability and dynamic availability at the control level of the hierarchy.

Software Defined Networking (SDN) has been a popular solution to deal with the heterogeneity of network domains, technologies, and vendors~\cite{Vilalta16, Baranda18}. 
Another example is the usage of LiFi-RF (Light Fidelity - Radio Frequency) heterogeneous networks~\cite{hansini1}. The additional capacity and low latency advocate for using different wireless technologies in the same indoor scenario. 
Multiple options to establish a connection improve reliability and remove the dependency on one technology. However, the sovereignty of such multi-domain networks still needs to be evaluated. 

\subsubsection{End-to-End (E2E) communication across multiple domains}


In multi-domain continental networks shown in Fig.\ref{fig:relation}, each country is a domain. Market fragmentation has led to several network and cloud/data center operators, each focused on different countries, making multi-domain services difficult and costly. For example, in~\cite{Bernados16}, a Europe-wide multi-domain platform is proposed for cross-domain orchestration. Turning more towards the reliability aspect, several works discuss survivability in multi-domain networks \cite{maria1}, based on link-disjoint~\cite{Gao14} or domain-disjoint~\cite{Gao11} inter-domain routing or geographically correlated failures~\cite{Riti18}. Though these ideas improve survivability and reliability, they do not account for network sovereignty. If some data or technology is legal in one domain but illegal in another, access to this data or service needs to be extensively monitored from both domains. Such a scenario seriously challenges establishing sovereign E2E communication in multi-domain networks.

Despite having multiple challenges, as a first step towards network sovereignty, in this work, we focus on the major challenge of developing a metric to measure network sovereignty mentioned in Sec.~\ref{sec:metriclack}. Through this metric, we believe that we can start addressing the other challenges by being able to compare the sovereignty of two different network configurations in different scenarios. 
Tackling the other challenges is noted as future work.

%% file: 4_Metric.tex
\section{Path Set Diversity metric}
\label{chap:metric}
This section discusses the proposed network sovereignty metric- Path Set Diversity (PSD) score. We start with guidelines to build a sovereign network and then define our metric.

\subsection{Guidelines for a sovereign network}
\label{sec:req}
A sovereign network must minimize the dependency on a certain manufacturer. Therefore, the basic guidelines for building a sovereign network are as follows.
\begin{enumerate}[label=({G}\arabic*),align=left]
	\item\label{statement:M1} Multiple paths between any source-destination pair.
	\item\label{statement:M2} Multiple manufacturers in the network.
	\item\label{statement:M3} Different manufacturer combinations in each path.
	\item\label{statement:M4} As few manufacturers as possible in each path.
\end{enumerate}

Note that these guidelines are all required to ensure maximum sovereignty. 

\begin{center}
\begin{figure}[tb]
\begin{subfigure}{0.48\linewidth}
\includegraphics[width=1\linewidth, center]{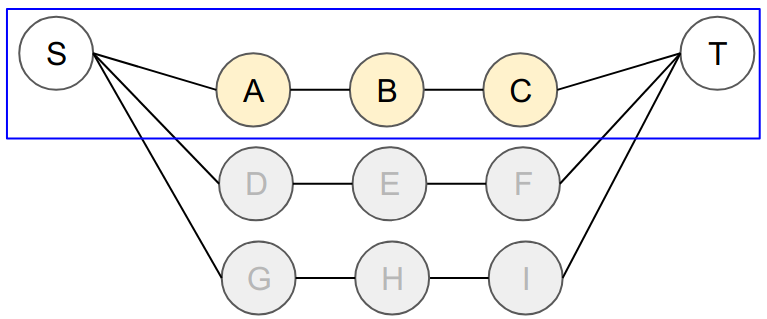} 
\caption{Single path between $S$ and $T$.}
\label{fig:M0}
\end{subfigure}
\begin{subfigure}{0.48\linewidth}
\includegraphics[width=1\linewidth, center]{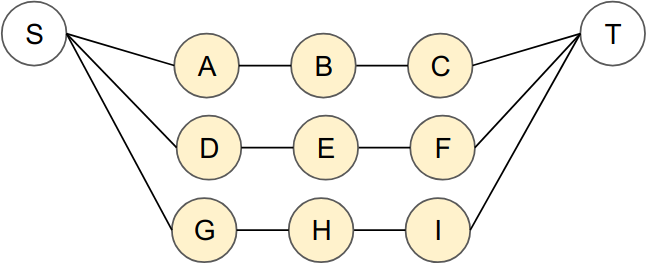} 
\caption{\ref{statement:M1} realized.}
\label{fig:M1}
\end{subfigure}
\begin{subfigure}{0.48\linewidth}
\includegraphics[width=1\linewidth, center]{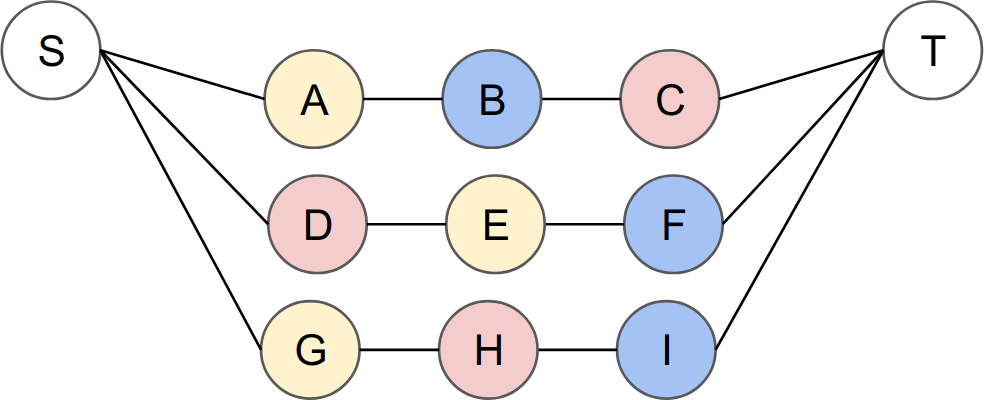} 
\caption{\ref{statement:M1} +~\ref{statement:M2} realized.}
\label{fig:M2}
\end{subfigure}
\begin{subfigure}{0.48\linewidth}
\includegraphics[width=1\linewidth, center]{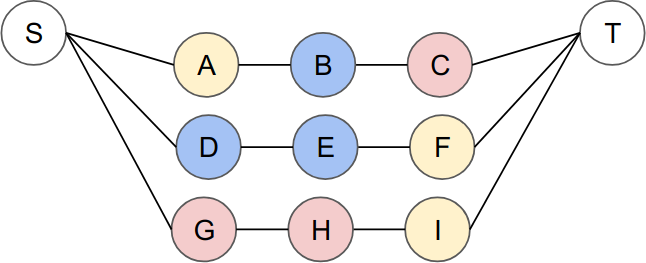} 
\caption{\ref{statement:M1} +~\ref{statement:M2} +~\ref{statement:M3} realized.}
\label{fig:M3}
\end{subfigure}
\begin{center}
\begin{subfigure}{0.6\linewidth}
\includegraphics[width=0.8\linewidth, center]{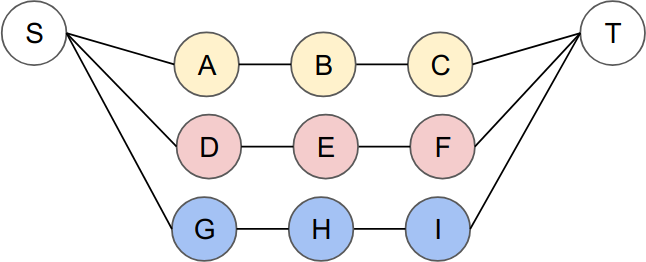} 
\caption{\ref{statement:M1} +~\ref{statement:M2} +~\ref{statement:M3} +~\ref{statement:M4} realized.}
\label{fig:M4}
\end{subfigure}
\end{center}
\caption{Illustrated guidelines for a sovereign network.}
\label{fig:requirements}
\end{figure}
\end{center}
Let us break down the list of guidelines with a simple progressive example shown in Fig.~\ref{fig:requirements}. 
First, consider the network in Fig.~\ref{fig:M0}. Consider a flow $r$ between the source $S$ and destination $T$. Let all the nodes in the path be from the yellow manufacturer. If there is any failure in the path, the flow $r$ fails. Therefore, we must follow the guideline~\ref{statement:M1} and have multiple paths between $S$ and $T$. 

Now, consider the network in Fig.~\ref{fig:M1}. Consider the same flow $r$ between the source $S$ and destination $T$, but now with several paths. Since all the nodes in all the paths are from the same yellow manufacturer, if the yellow manufacturer fails, all the nodes in the paths fail. Therefore, we need to follow guideline~\ref{statement:M2} and have multiple manufacturers in the network to avoid a dependency on a single manufacturer.

Next, consider the network in Fig.~\ref{fig:M2}. Consider the same flow $r$ between the source $S$ and destination $T$. The nodes in the paths are now purchased from three different manufacturers- yellow, blue, and red. However, since all the paths have the same manufacturer combination with all manufacturers present in each path, any one manufacturer failing will cause the flow $r$ to fail. Hence, we must follow the guideline~\ref{statement:M3} and have different manufacturer combinations. Note that the first path, S-A-B-C-T, and the second path, S-D-E-F-T, have different manufacturers' orders. However, they are still considered the same manufacturer combination from a sovereignty perspective.

Next, consider the network in Fig.~\ref{fig:M3}. Consider the same flow $r$ between the source $S$ and destination $T$. The nodes in the paths are now from different manufacturers such that the manufacturer combinations in each path are different. This case is better than the previous cases because if the red manufacturer or blue manufacturer fails separately, the connectivity between $S$ and $T$ still remains intact. Yet, the failure of the yellow manufacturer will cause flow $r$ to fail. Therefore, we must add guideline~\ref{statement:M4} and have as few manufacturers as possible in each path.

Finally, consider the network in Fig.~\ref{fig:M4}. Consider the same flow $r$ between the source $S$ and destination $T$. Now, all the four guidelines~\ref{statement:M1},~\ref{statement:M2},~\ref{statement:M3}, and~\ref{statement:M4} are followed. 
Guideline~\ref{statement:M4} is the least intuitive guideline. Having fewer manufacturers in each path means fewer dependencies in each path. Therefore, in this example, when any manufacturer fails, at least two paths exist between $S$ and $T$. Furthermore, this arrangement can even tolerate two manufacturers failing simultaneously. This manufacturer assignment can now be called sovereign, as it removed the dependency on manufacturers.

Therefore, quantitatively, network sovereignty can be defined as the ability to operate a network acceptably satisfying the requirements, even when all but one manufacturer fails.
In a real-world scenario like the multi-vendor disaggregated network discussed in Section~\ref{sec:disaggregated_network}, a manufacturer assignment based on the guidelines~\ref{statement:M1},~\ref{statement:M2},~\ref{statement:M3}, and~\ref{statement:M4} can guarantee maximum sovereignty.

Note that there could be multiple manufacturer assignments that can guarantee maximum network sovereignty. 
Moreover, we still have not quantified how sovereign the arrangements in Fig.~\ref{fig:requirements} are.
We aim to develop a metric that captures the sovereignty of the network's manufacturer assignment by introducing the PSD score metric in Section~\ref{sec:psdmetric}.

\subsection{Path Set Diversity (PSD) score}
\label{sec:psdmetric}
PSD in a network is defined as the extent of manufacturer diversity in the distinct paths used for each flow.
PSD score awards the network when the guidelines in Section~\ref{sec:req} are met. 
There are four points to consider.
\begin{enumerate}[label=({P}\arabic*),align=left]
\item Irrespective of the number of nodes in a path, the number of manufacturers in the path is relevant for the PSD score because, if a manufacturer fails, all nodes from that manufacturer fail.
\item If two paths with the same manufacturer combination exist, they do not add any value to the PSD score. This is because the same failure can cause both paths to fail simultaneously. Therefore, the redundant paths with the same manufacturer combination must be removed. 
\item \label{statement: P3} The order of manufacturers in the combination also does not matter, i.e., a combination like $\{$manufacturer-0, manufacturer-1$\}$ is the same as $\{$manufacturer-1, manufacturer-0$\}$. The combinations are commutative.
\item The number of distinct, simple paths between a source and a destination can be very high for a large network. 
However, not all the paths can be used to route traffic. 
For example, some paths may be too long. Hence, they can have high latency or high costs. 
To account for this, $k$-shortest distinct paths are counted in our work. $k$ is an operator choice and can be chosen based on several factors.
For example, the operator may consider the paths for which the total latency is below a particular threshold. $k$ can also be chosen based on the cost of operating a particular link. $k$ is a design parameter, and its impact on the results is discussed later in Section~\ref{sec:impactK}.
\end{enumerate}


Each flow in the network is evaluated separately based on the manufacturer assignments in its $k$ paths. Each path is assigned a reward called path reward, $R_p$. Each flow is assigned a reward called flow reward, $\pi_r$. The proposed network sovereignty metric, the PSD score denoted by $\Pi$, is numerically evaluated as the weighted average of all the flow rewards $\pi_r$. The evaluation of $R_p$, $\pi_r$, and $\Pi$ is further explained in this section.
The Algorithm~\ref{algo:psd} describes calculating the PSD score for a manufacturer-assigned network.

\begin{algorithm}
\caption{Path Set Diversity score calculation.}
\label{algo:psd}
\SetAlgoLined
\DontPrintSemicolon
\ForEach{flow ($r = (s_r, t_r$))}
{
Find the $k$-shortest paths from source ($s_r$) to destination ($t_r$)\;
	\ForEach{path}
	{
    	Find the manufacturers in the path ($m_p$).\;
    }
    Remove redundant paths (the ones with the same combination of manufacturers).\;
	\ForEach{path}
	{Find the path reward $R_p = \frac{1}{|m_p|}$. \;
    }
	Find the flow reward  $\pi_r = \sum\limits_{p \in k}R_p$. \;      		
}
\label{algo}
\end{algorithm}

\begin{figure}
\centering
\includegraphics[width=0.45\linewidth]{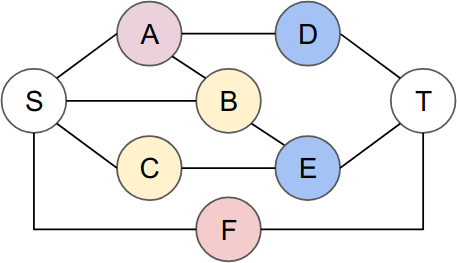}
\caption{Example network for PSD score calculation.}
\label{fig:psdexample}
\end{figure}

Consider the sample network shown in Fig.~\ref{fig:psdexample} to explain the PSD score calculation. Each flow $r$ in the network is represented by its source ($s_r$) and destination ($t_r$) as $r = (s_r, t_r)$. Let us consider a flow $r$ with source-destination pair $S, T$ in the network in Fig.~\ref{fig:psdexample}. Let the colors yellow (Y), blue (B), and red (R) denote three different manufacturers. Table~\ref{tab:calc} shows the PSD score calculation. First, the `Path' column notes the $k$-shortest paths from $S$ to $T$, according to Line~2 of Algorithm~\ref{algo:psd}. Then, the manufacturers in each path are identified in the next column, as per Lines~3-5 of Algorithm~\ref{algo:psd}. Then, paths 4, 6, and 7 are removed as per Line~6 of Algorithm~\ref{algo:psd} because path 4 is redundant to path 3, while paths 6 and 7 are redundant to path 5. As discussed in point~\ref{statement: P3}, the order in $m_p$ does not matter. Lines 5, 6, and 7 have the same manufacturer combination in different orders. Therefore, only one iteration can be counted.
After removing redundant paths, the number of manufacturers ($|m_p|$) in each path is counted in the following column. The path reward ($R_p$) is calculated as the inverse of $|m_p|$ in the last column as per Lines~7-10 of Algorithm~\ref{algo:psd}. 
The inverse of $|m_p|$ is used to reward the path with lower $|m_p|$, and penalize the path with higher $|m_p|$. This ensures that the fourth guideline~\ref{statement:M4} is considered in the PSD score.
Finally, the flow reward $\pi_r$ for the flow $r$ is obtained by summing up all the $R_p$ as per Line~11 of Algorithm~\ref{algo:psd}. In this example, $\pi_r$ is calculated to be $2.33$.

In the best case scenario, the upper theoretical bound of the flow reward ($\widehat{\pi_r}$) is possible when every manufacturer combination exists in the $k$-shortest paths.
For three manufacturers, 
\begin{equation}
\begin{aligned}
\widehat{\pi_r} = &(3 \times \texttt{single manufacturer}) + \\ &(3 \times \texttt{two manufacturer combinations}) + \\ &(1 \times \texttt{all three manufacturers}), \\
\widehat{\pi_r} &= (3 \times 1) + (3 \times 0.5) + (1 \times 0.33) = 4.83. 
\end{aligned}
\end{equation}
For easier calculation, the upper theoretical bound for the flow reward ($\widehat{\pi_r}$) for $|M|$ number of manufacturers is given by,
\begin{equation}
\widehat{\pi_r} = \sum\limits_{i=0}^{|M|-1} \Biggl( \binom{|M|}{i} \cdot \frac{1}{i} \Biggr)
\end{equation}
However, this optimal flow reward value is difficult to achieve in a real-world deployment because the nodes used by this flow may be shared by other flows, which may prefer a different manufacturer assignment. 
Therefore, network operators must aim to get their flow rewards as close to $\widehat{\pi_r}$ as possible.


The proposed network sovereignty metric, the PSD score ($\Pi$), is mathematically computed as the weighted average of all the flow rewards as seen in Eq.~\ref{eq:psd}, where $w_r$ is the weight of flow $r$. The flow weights ensure that more priority is given to a more important flow. The operator may award weights based on factors like flow size, flow priority, etc.

\begin{equation}
\label{eq:psd}
\Pi = \frac{\sum\limits_{r \in R} \bigl(w_r \times \pi_r \bigr)}{\sum\limits_{r \in R} w_r}
\end{equation}

Interestingly, considering the ratio of achieved $\pi_r$ to the upper theoretical bound $\widehat{\pi_r}$ is incorrect because there will be no difference in the PSD score for having two or four manufacturers when the PSD score is given as a percentage value. 
Guideline~\ref{statement:M2} states that more manufacturers are needed in the network, and this will not be accounted for in the PSD score if the percentage value is considered. 


From Eq.~\ref{eq:psd}, we have proposed a metric to quantify the sovereignty of a network using the PSD score, $\Pi$. Now, this $\Pi$ can be used to compare the sovereignty of different manufacturer assignments in the same network and the sovereignty of different networks that are similar in terms of the number of nodes, number of edges, number of $k$-shortest paths, and number of hops in the shortest paths. Only similarly sized networks can be compared fairly because the PSD score is directly dependent on the aforementioned input graph attributes, similar to the other dependability attributes.

\begin{table}[]
\caption{Example for PSD score calculation.}
\label{tab:calc}
\begin{center}
\begin{tabular}{|ccccc|}
\hline
\multicolumn{1}{|c|}{\multirow{2}{*}{\begin{tabular}[c]{@{}c@{}}Path\\ no.\end{tabular}}} &
  \multicolumn{1}{c|}{\multirow{2}{*}{Path}} &
  \multicolumn{2}{c|}{\begin{tabular}[c]{@{}c@{}}Manufacturers\\ in Path\end{tabular}} &
  \multirow{2}{*}{$R_p = \frac{1}{|m_p|}$} \\ \cline{3-4}
\multicolumn{1}{|c|}{} &
  \multicolumn{1}{c|}{} &
  \multicolumn{1}{c|}{$m_p$} &
  \multicolumn{1}{c|}{$|m_p|$} &
   \\ \hline
\multicolumn{1}{|c|}{1} &
  \multicolumn{1}{c|}{S-F-T} &
  \multicolumn{1}{c|}{R} &
  \multicolumn{1}{c|}{1} &
  1/1 \\ \hline
\multicolumn{1}{|c|}{2} &
  \multicolumn{1}{c|}{S-A-D-T} &
  \multicolumn{1}{c|}{R,B} &
  \multicolumn{1}{c|}{2} &
  1/2 \\ \hline
\multicolumn{1}{|c|}{3} &
  \multicolumn{1}{c|}{S-B-E-T} &
  \multicolumn{1}{c|}{Y,B} &
  \multicolumn{1}{c|}{2} &
  1/2 \\ \hline
\multicolumn{1}{|c|}{4} &
  \multicolumn{1}{c|}{S-C-E-T} &
  \multicolumn{1}{c|}{Y,B} &
  \multicolumn{2}{c|}{\begin{tabular}[c]{@{}c@{}}Removed, because redundant\\ to Path 3\end{tabular}} \\ \hline
\multicolumn{1}{|c|}{5} &
  \multicolumn{1}{c|}{S-A-B-E-T} &
  \multicolumn{1}{c|}{R,Y,B} &
  \multicolumn{1}{c|}{3} &
  1/3 \\ \hline
\multicolumn{1}{|c|}{6} &
  \multicolumn{1}{c|}{S-B-A-D-T} &
  \multicolumn{1}{c|}{Y,R,B} &
  \multicolumn{2}{c|}{\begin{tabular}[c]{@{}c@{}}Removed, because redundant\\ to Path 5\end{tabular}} \\ \hline
\multicolumn{1}{|c|}{7} &
  \multicolumn{1}{c|}{S-C-E-B-A-D-T} &
  \multicolumn{1}{c|}{Y,R,B} &
  \multicolumn{2}{c|}{\begin{tabular}[c]{@{}c@{}}Removed, because redundant\\ to Path 5\end{tabular}} \\ \hline
\multicolumn{4}{|r|}{$\pi_r$} &
  2.33 \\ \hline
\end{tabular}
\end{center}
\end{table}

%% file: 5_Naga.tex
\section{Manufacturer Assignment for Sovereignty (MAS) and \textit{Naga}}
\label{chap:Naga}

A network operator must plan how many network components to buy, how many manufacturers to buy from, and how to place these components from different vendors in the network topology
From~\cite{mine1, mine2}, appropriately arranging nodes from different manufacturers is crucial in removing dependencies on vendors and thereby improving network sovereignty. Working on this finding, we introduce the Manufacturer Assignment for Sovereignty (MAS) problem in this section and propose \textit{Naga}, our ILP solution to maximize the PSD score $\Pi$, and thereby the network sovereignty.

Given a network, the expected traffic, the number of manufacturers ($|M|$) available, and the $k$-shortest paths chosen by the operator, the MAS problem condenses to the following question.
\begin{enumerate}[label=({Q}\arabic*),align=left]
	\item\label{q:Q1} What is the best manufacturer assignment possible to maximize network sovereignty?
\end{enumerate}

\textit{Naga} is our proposed ILP solution that solves the MAS problem by optimizing for the PSD score, $\Pi$.
This section describes the formulation of \textit{Naga}.

\subsection{Constants}
\textit{Naga}'s formulation considers the following constants.
\begin{itemize}
\item Let us consider a network represented as a graph $G$ with vertices $V$ and edges $E$. 
\item Let $M$ be the set of manufacturers that the operator can choose from. 
\item Let $|M|$ be the number of manufacturers. 
\item Let $R$ be the set of all the flows in the network. 
\item Each flow $r \in R$ is represented by its source ($s_r$) and destination ($t_r$) as $r = (s_r, t_r)$. 
\item Let $k$ be the number of shortest paths the operator considers for each flow. 
\item Let the list of $k$-shortest paths for flow $r$ be $P_r = \{p_{r0}, p_{r1}, ... p_{r(k-1)} \}$. If there are less than $k$-shortest paths possible due to the low redundancy in the network, only the possible paths are counted for that flow.
\item Let the set of nodes in the $j^{th}$ path of flow $r$ be $p_{rj}$ such that $p_{rj}$ does not include the source and destination nodes. In PSD score calculation, the source and destination nodes are neglected, similar to~\cite{mine1, mine2}, because if the source and destination nodes fail, the flow fails irrespective of the network's sovereignty.
Hence, $P_r$ does not have any one-hop paths.
\item The manufacturer combination $x$ in the path governs the PSD score. $x_m$ is the binary notation equal to one if manufacturer $m$ is present in $x$.
The manufacturer combination is represented as, $x = \{x_m | x_m \in \mathbb{Z}_{2}, m \in M\}$. For example, if $|M| = 3$, and if the path has the second and third manufacturers only, irrespective of the number of nodes in the path, then $x_0 = 0, x_1 = 1, x_2 = 1$, and hence, $x = 011$.
\item The list of all possible manufacturer combinations is given by $X = \{x | x \in \mathbb{Z}_{2}^{|M|}\}$ subject to $x \neq (000)$ because at least one manufacturer is required. For example, if $|M| = 3$, then $X = \{001, 010, 100, 011, 110, 101, 111\}$.
\item PSD score rewards the network when multiple combinations of manufacturers are possible. This reward $q_x$ for a manufacturer combination $x$ equals the inverse of the number of manufacturers present in the combination as discussed in Table~\ref{tab:calc} in Section~\ref{sec:psdmetric}. The number of manufacturers in combination $x$ is essentially the number of ones in $x$. Therefore, $q_x$ can be represented as $\frac{1}{\sum\limits_{m \in M} x_m}$.
\end{itemize}
Table~\ref{tab:constants} summarizes the constants. 
Since $x$ denotes the manufacturer combination in a path, it is equivalent to $m_p$ from Table~\ref{tab:calc} in Section~\ref{sec:psdmetric}.
Similarly, the number of ones in $x$ denoting the number of manufacturers present in the path is equivalent to $|m_p|$ from Table~\ref{tab:calc} in Section~\ref{sec:psdmetric}. Since the ILP requires the binary representation of the manufacturers in a path, we use the notation $x$ instead of $m_p$ in this section.

\begin{table}[]
\centering
\caption{\textit{Naga}- Constants.}
\label{tab:constants}
\begin{tabular}{|p{3.8cm}|p{4cm}|}
\hline
\textbf{Constants}                      & \textbf{Symbols}      \\ \hline 
Graph             & $G = (V,E)$              \\ \hline
Set of edges                            & $E$                   \\ \hline
Set of vertices                       & $V$                   \\ \hline 
Set of manufacturers & $M = \{0,1,2, .. |M|-1\}$ \\ \hline
No. of manufacturers & $|M|$ \\ \hline
Flow (src,dst)   & $r = (s_r, t_r)$ \\ \hline
Set of flows & $R$ \\ \hline
No. of shortest paths considered for each flow & $k$ \\ \hline
Set of possible paths for flow $r$ & $P_r = \{p_{r0}, p_{r1}, ... p_{r(k-1)} \}$ \\ \hline
Set of nodes in the $j^{th}$ path of flow $r$ & $(s_r, t_r) \notin p_{rj} $ \\ \hline
Binary notation equal to one if manufacturer $m$ is present in the manufacturer combination & $x_m$ \\ \hline
Possible manufacturer combination & $x = \{x_m | x_m \in \mathbb{Z}_{2}, m \in M\}$ \\ \hline
All possible manufacturer combinations & $X = \{x | x \in \mathbb{Z}_{2}^{|M|} \land x \neq (000)\}$\\ \hline
PSD reward for a combination $x$ & $q_x = \frac{1}{\sum \text{Ones in } x}  = \frac{1}{\sum\limits_{m \in M} x_m} $ \\ \hline
\end{tabular}
\end{table}

\subsection{Variables}
\label{sec:subvariable}
\textit{Naga}, in the simplest of words, is a manufacturer assignment solution. Therefore, the most important binary decision variable $b_{mn}$ denotes if node $n$ is bought from manufacturer $m$.
For the PSD score calculation, we need to identify which manufacturer is used in which path for which flow. Hence, $u_{mrj}$ is the binary variable equal to one if manufacturer $m$ is used in the $j^{th}$ path of flow $r$. 
Furthermore, the binary variable $f_{xrj}$ is equal to 1 if combination $x$ is used in the $j^{th}$ path of flow $r$.
Another binary variable $F_{xr}$ is also needed to know if the combination $x$ is used in flow $r$. This is specifically included to enforce Line~8 in Algorithm~\ref{algo:psd}. For example, let a flow $r$ have two paths, $p_{r1}$ and $p_{r2}$, with the same manufacturer assignment, say $\hat{x}$. In that case, $f_{\hat{x}r1}$ and $f_{\hat{x}r2}$ are both equal to one. In such a case, the reward $q_{\hat{x}}$ for the combination $\hat{x}$ should not be counted twice but once. Therefore, the variable $F_{xr}$ removes this redundancy in this case by being $F_{\hat{x}r} = 1$.
Furthermore, the accumulated PSD reward for flow $r$ is given by $\pi_r$.
Table~\ref{tab:variables} summarizes the variables and their symbols.

\begin{table}[]
\centering
\caption{\textit{Naga}- Variables.}
\label{tab:variables}
\begin{tabular}{|p{6.5cm}|p{1cm}|}
\hline
\textbf{Variables}                      & \textbf{Symbols}      \\ \hline 
Binary variable equal to one if node $n$ is from manufacturer $m$             & $b_{mn}$            \\ \hline
Binary variable equal to one if manufacturer $m$ is used in the $j^{th}$ path of flow $r$       & $u_{mrj}$     \\ \hline
Binary variable equal to one if combination $x$ is used in the $j^{th}$ path of flow $r$       & $f_{xrj}$     \\ \hline
Binary variable equal to one if combination $x$ is used in flow $r$       & $F_{xr}$     \\ \hline
Accumulated PSD reward for flow $r$      & $\pi_r$     \\ \hline
\end{tabular}
\end{table}
\subsection{Constraints}
The first constraint in Eq.~\ref{eq:constraint1} is to restrict the purchase of one node from one manufacturer only.
\begin{equation}
\label{eq:constraint1}
\sum\limits_{m \in M} b_{mn} = 1 , \forall n \in V
\end{equation}
Next, the variable $u_{mrj}$ must be forced to one if manufacturer $m$ is used in the $j^{th}$ path of flow $r$ as per Eq.~\ref{eq:constraint2}. This is achieved by the binary `or' $\bigvee$ operator.
\begin{equation}
\label{eq:constraint2}
u_{mrj} = \bigvee\limits_{n \in p_{rj}} b_{mn} , \forall m \in M, \forall r \in R, \forall p_{rj} \in P_r
\end{equation}
Additionally, the variable $f_{xrj}$ must be forced to one if combination $x$ is used in the $j^{th}$ path of flow $r$  as per Eq.~\ref{eq:constraint3}. This is achieved by the binary `and' $\bigwedge$ operator. All possible binary combinations of $x \in X$ are evaluated.
\begin{equation}
\label{eq:constraint3}
f_{xrj} = \bigwedge\limits_{x_m \in x}(\widetilde{u_{mrj}}) , \forall r \in R, \forall p_{rj} \in P_r, \forall x \in X,
\end{equation}
\begin{equation}
 \text{where, } \widetilde{u_{mrj}} = \begin{cases}
      u_{mrj} & \text{if $x_m = 1$}\\
      1-u_{mrj}& \text{if $x_m = 0$}
    \end{cases}
\end{equation}
For example, if $|M| = 3$, and if the $j^{th}$ path of flow $r$ has the second and third manufacturers only, then for $x = 011$,
\begin{equation}
f_{xrj} = f_{(011)rj} = (1 - u_{0rj}) \land u_{1rj} \land u_{2rj} = 1.
\end{equation}
Another constraint is needed to remove the duplicated redundancies in the PSD score calculation as stated in Line~6 of Algorithm~\ref{algo:psd}. As discussed in Section~\ref{sec:subvariable}, $F_{xr}$ is forced to 1 if combination $x$ is used in at least one path of flow $r$ by using the binary `or' operator for each flow over all possible manufacturer combinations $x \in X$ as shown in Eq.~\ref{eq:constraint4}.
\begin{equation}
\label{eq:constraint4}
F_{xr} = \bigvee\limits_{\forall p_{rj} \in P_r} f_{xrj}, \forall x \in X , \forall r \in R
\end{equation}
Finally, the binary variable $F_{xr}$ is weighted with its corresponding reward $q_x$ for each combination $x$ to obtain the accumulated PSD reward for a flow $r$ as shown in Eq.~\ref{eq:constraint5}.
\begin{equation}
\label{eq:constraint5}
\pi_r = \sum\limits_{x \in X} (q_x \times F_{xr}), \forall r \in R
\end{equation}
Note that the binary `or' $\bigvee$ and binary `and' $\bigwedge$ operators can be directly implemented as part of the program formulation on optimization suites like Gurobi \cite{gurobi}.

\subsection{Objective}
\label{sec:objective}
The goal of \textit{Naga} is to solve the MAS problem by finding the best manufacturer assignment to maximize network sovereignty.
To achieve this, \textit{Naga} aims to maximize the PSD score. Therefore, the mathematical representation of \textit{Naga}'s objective is maximizing Eq.~\ref{eq:psd}, represented as,
\begin{equation}
\label{eq:o1}
max \Biggl( \frac{ \sum\limits_{r \in R} ( w_r \times \pi_r) } {\sum\limits_{r \in R} w_r} \Biggr),
\end{equation}
where $w_r$ is the weight assigned to each flow.
Note that $q_x$ in Eq.~\ref{eq:constraint5} and $w_r$ in Eq.~\ref{eq:o1} are constants, and they do not make \textit{Naga}'s formulation non-linear.

%% file: 5b_evalworkflow.tex
\section{Evaluation workflow}
\label{chap:evalworkflow}
\begin{figure*}[tb]
\begin{subfigure}{0.24\linewidth}
    \includegraphics[width=1\linewidth]{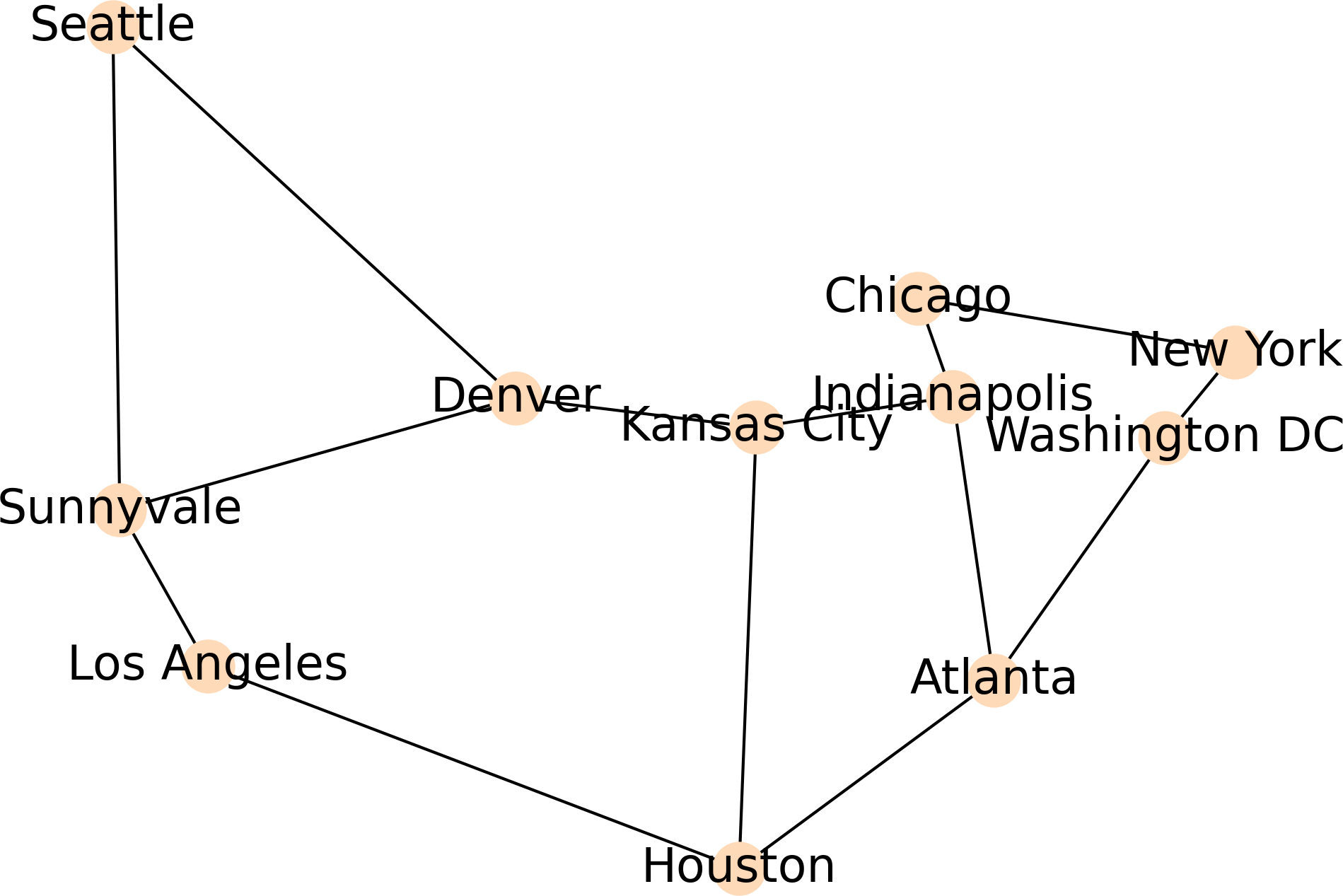}
    \caption{Abilene.} 
    \label{fig:abilene}
\end{subfigure}
\begin{subfigure}{0.24\linewidth}
    \includegraphics[width=1\linewidth]{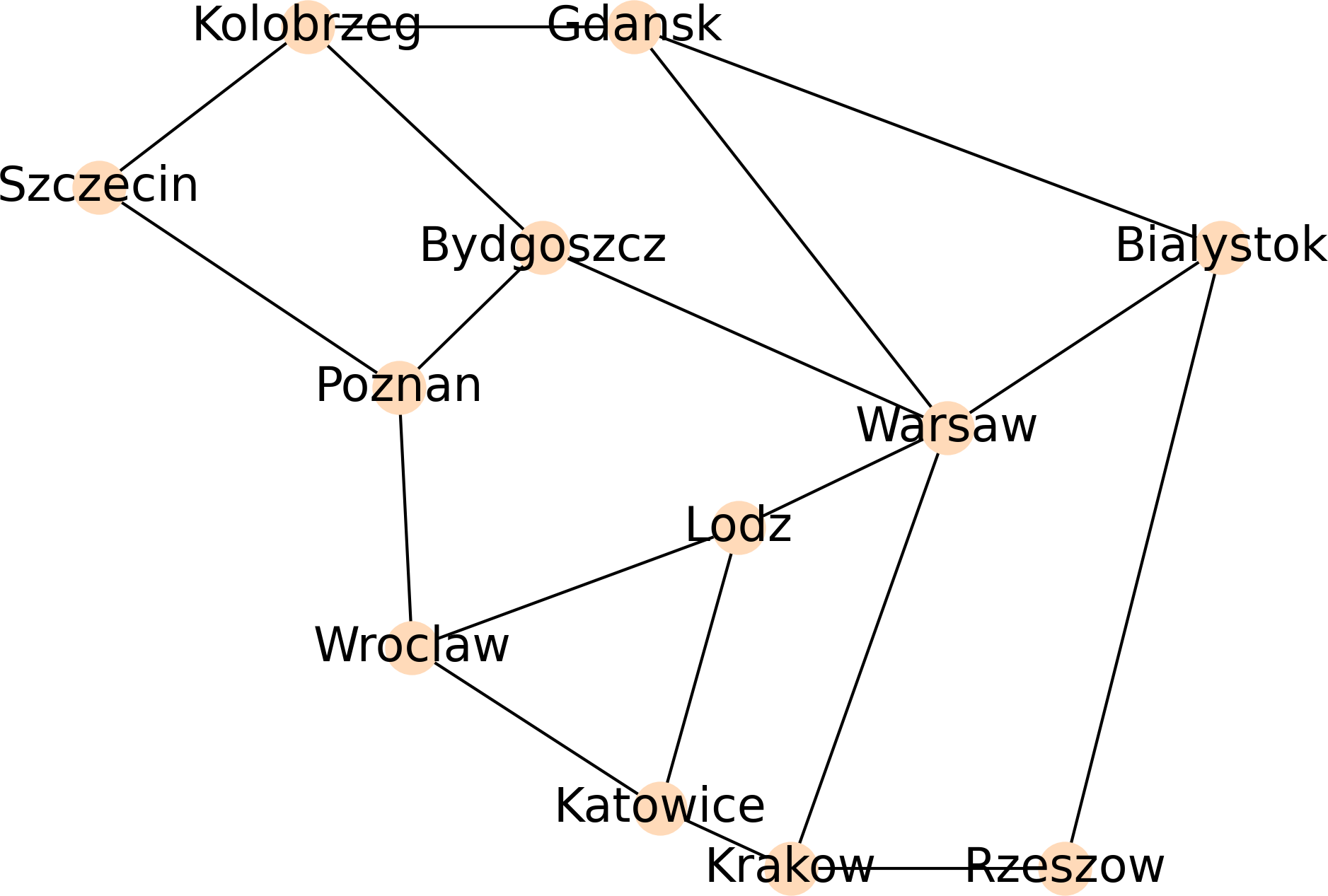}
    \caption{Polska.}  
    \label{fig:polska}
\end{subfigure}
\begin{subfigure}{0.24\linewidth}
    \includegraphics[width=1\linewidth]{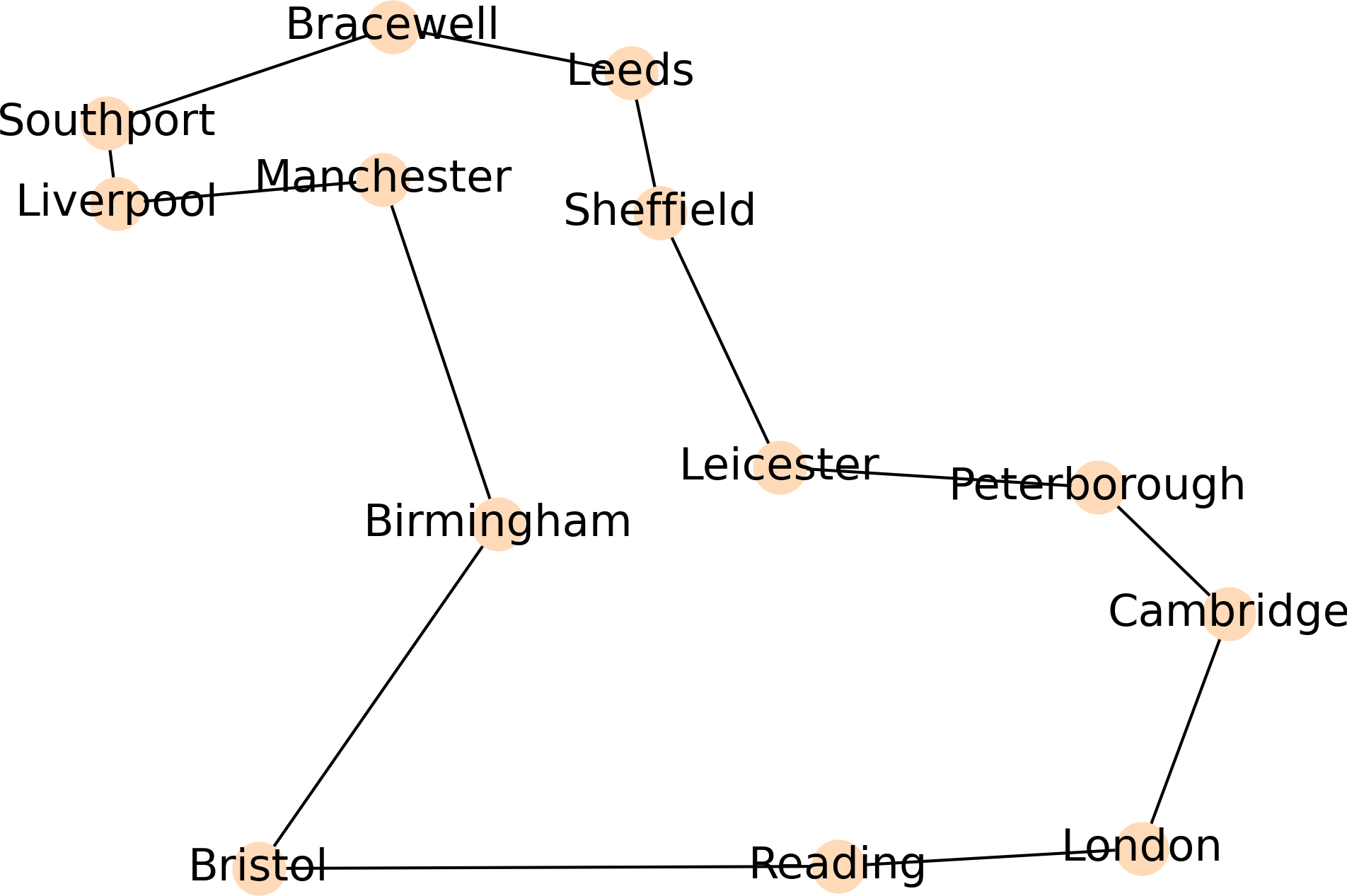}
    \caption{HiberniaUK.}
    \label{fig:uk}
\end{subfigure}
\begin{subfigure}{0.24\linewidth} 
    \includegraphics[width=1\linewidth]{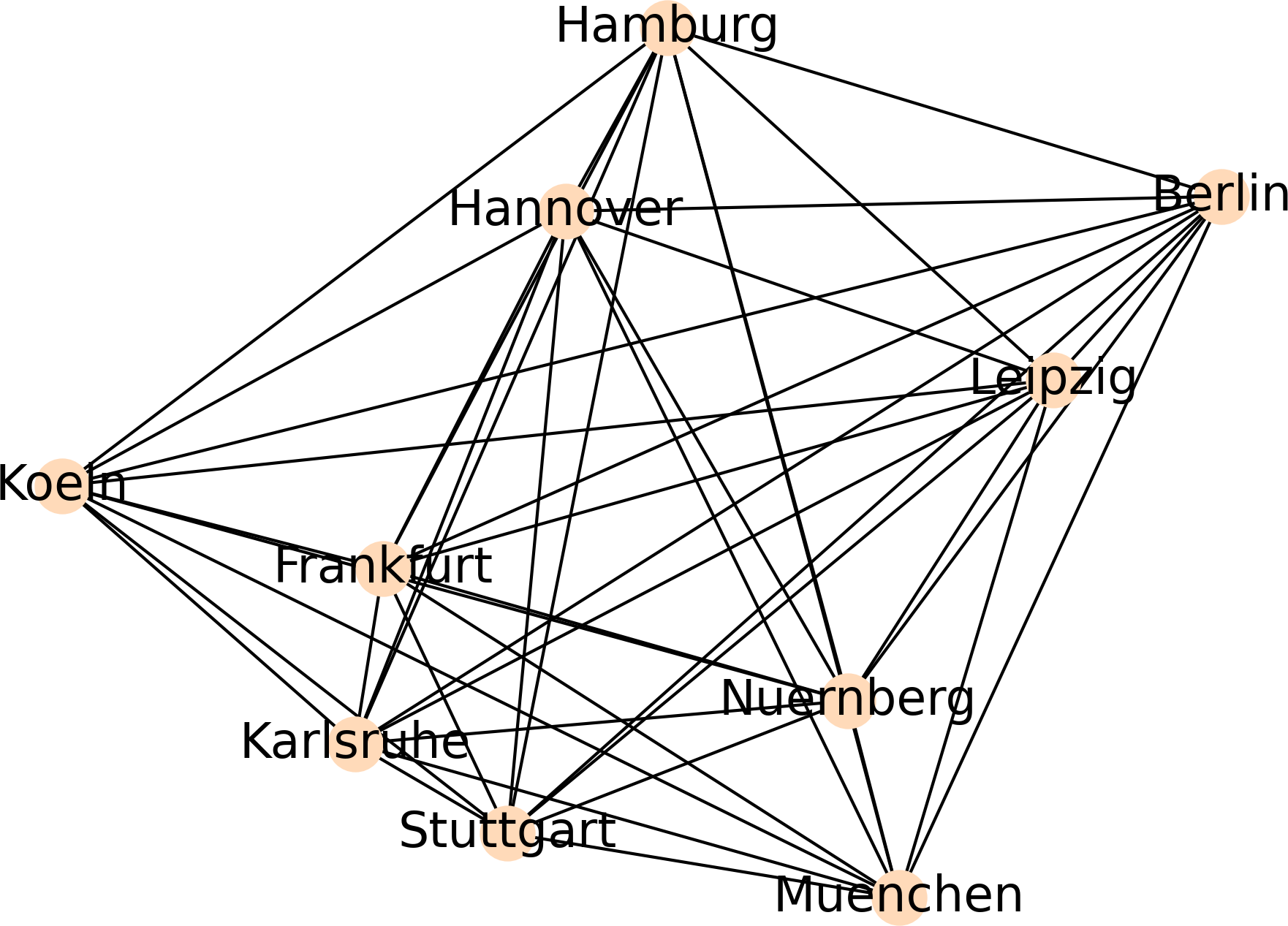}
    \caption{dfn-bwin.}
    \label{fig:de}
\end{subfigure}
\caption{Topologies considered in this study.}
\label{fig:topos}
\end{figure*}

This section evaluates and discusses the results of \textit{Naga}. To test the performance of the PSD score metric and \textit{Naga}, we evaluate both on four real-world core networks as shown in Table~\ref{tab:ip} and in Fig.~\ref{fig:topos}. First, we discuss the input parameters.

\subsection{Input parameters}
\label{sec:inputpara}

We consider four topologies- Abilene, Polska, HiberniaUK, and dfn-bwin. HiberniaUK is a ring topology, while the dfn-bwin topology is a full mesh. We consider these two to see \textit{Naga}'s performance in special cases. Although core networks are considered in this evaluation, the PSD score metric and \textit{Naga} can be applied to any other network. The simulation setup also considers various numbers of manufacturers ($|M| \in \{2,3,4,5\}$). This analysis will show the impact of using more manufacturers on network sovereignty. For each topology, any-to-any traffic is considered between all possible source-destination pairs.

\begin{table}[]
\centering
\caption{Input options.}
\label{tab:ip}
\begin{tabular}{|c|ccccc|}
\hline
{\color[HTML]{131313} \textbf{Input parameter}} &
  \multicolumn{5}{c|}{{\color[HTML]{131313} \textbf{Choices}}} \\ \hline
{\color[HTML]{131313} } &
  \multicolumn{1}{l|}{} &
  \multicolumn{1}{c|}{{\color[HTML]{131313} $|V|$}} &
  \multicolumn{1}{c|}{{\color[HTML]{131313} $|E|$}} &
  \multicolumn{1}{c|}{{\color[HTML]{131313} Source}} &
  {\color[HTML]{131313} Figure} \\ \cline{2-6} 
{\color[HTML]{131313} } &
  \multicolumn{1}{c|}{{\color[HTML]{131313} Abilene}} &
  \multicolumn{1}{c|}{{\color[HTML]{131313} 11}} &
  \multicolumn{1}{c|}{{\color[HTML]{131313} 14}} &
  \multicolumn{1}{c|}{{\color[HTML]{131313}~\cite{zoo}}} &
  {\color[HTML]{131313}~\ref{fig:abilene}} \\ \cline{2-6} 
{\color[HTML]{131313} } &
  \multicolumn{1}{c|}{{\color[HTML]{131313} Polska}} &
  \multicolumn{1}{c|}{{\color[HTML]{131313} 12}} &
  \multicolumn{1}{c|}{{\color[HTML]{131313} 18}} &
  \multicolumn{1}{c|}{{\color[HTML]{131313}~\cite{sndlib}}} &
  {\color[HTML]{131313}~\ref{fig:polska}} \\ \cline{2-6} 
{\color[HTML]{131313} } &
  \multicolumn{1}{c|}{{\color[HTML]{131313} HiberniaUK}} &
  \multicolumn{1}{c|}{{\color[HTML]{131313} 13}} &
  \multicolumn{1}{c|}{{\color[HTML]{131313} 13}} &
  \multicolumn{1}{c|}{{\color[HTML]{131313}~\cite{zoo}}} &
  {\color[HTML]{131313}~\ref{fig:uk}} \\ \cline{2-6} 
\multirow{-5}{*}{{\color[HTML]{131313} Topology}} &
  \multicolumn{1}{c|}{{\color[HTML]{131313} dfn-bwin}} &
  \multicolumn{1}{c|}{{\color[HTML]{131313} 10}} &
  \multicolumn{1}{c|}{{\color[HTML]{131313} 45}} &
  \multicolumn{1}{c|}{{\color[HTML]{131313}~\cite{sndlib}}} &
  {\color[HTML]{131313}~\ref{fig:de}} \\ \hline
{\color[HTML]{131313} $ |M|$} &
  \multicolumn{5}{c|}{{\color[HTML]{131313} 2,3,4,5}} \\ \hline
\end{tabular}
\end{table}

For this preliminary study, we consider flows with equal weights. Therefore, $w_r = 1$, for all $r \in R$.
Therefore the objective in Eq.~\ref{eq:o1} discussed in Section~\ref{sec:objective} is rewritten as,
\begin{equation}
\label{eq:o2}
max \Bigl( \sum\limits_{r \in R} \pi_r \Bigr)
\end{equation}
Fewer communicating pairs will improve \textit{Naga}'s performance and decrease the run time because there are fewer common nodes in the paths for different communicating pairs. However, the ILP formulation remains the same. \textit{Naga}'s run time will most likely marginally increase with different flow weights.
All traffic is assumed to be bidirectional.

\subsection{Simulation setup and run time}
\label{sec:simsetup}
After setting up the topology and traffic, \textit{Naga} runs on Python with the Gurobi optimization suite~\cite{gurobi}. The program was run on an AMD Ryzen 3700X octa-core processor with 32GB RAM. \textit{Naga}'s run time for one topology, for two manufacturers ($|M| = 2$), was around 15-20 minutes, depending on the topology. With an increase in the number of manufacturers, the run time also increased due to the corresponding increase in variables and constraints. For example, the run time increased for three manufacturers ($|M| = 3$) to 30-35 minutes. The centrality metric-based heuristics run in the order of 10 seconds irrespective of $|M|$. However, the increased run time for \textit{Naga} is not a problem in a long-term network planning problem.
\begin{figure}[tb]
\begin{center}
\includegraphics[width=1\linewidth]{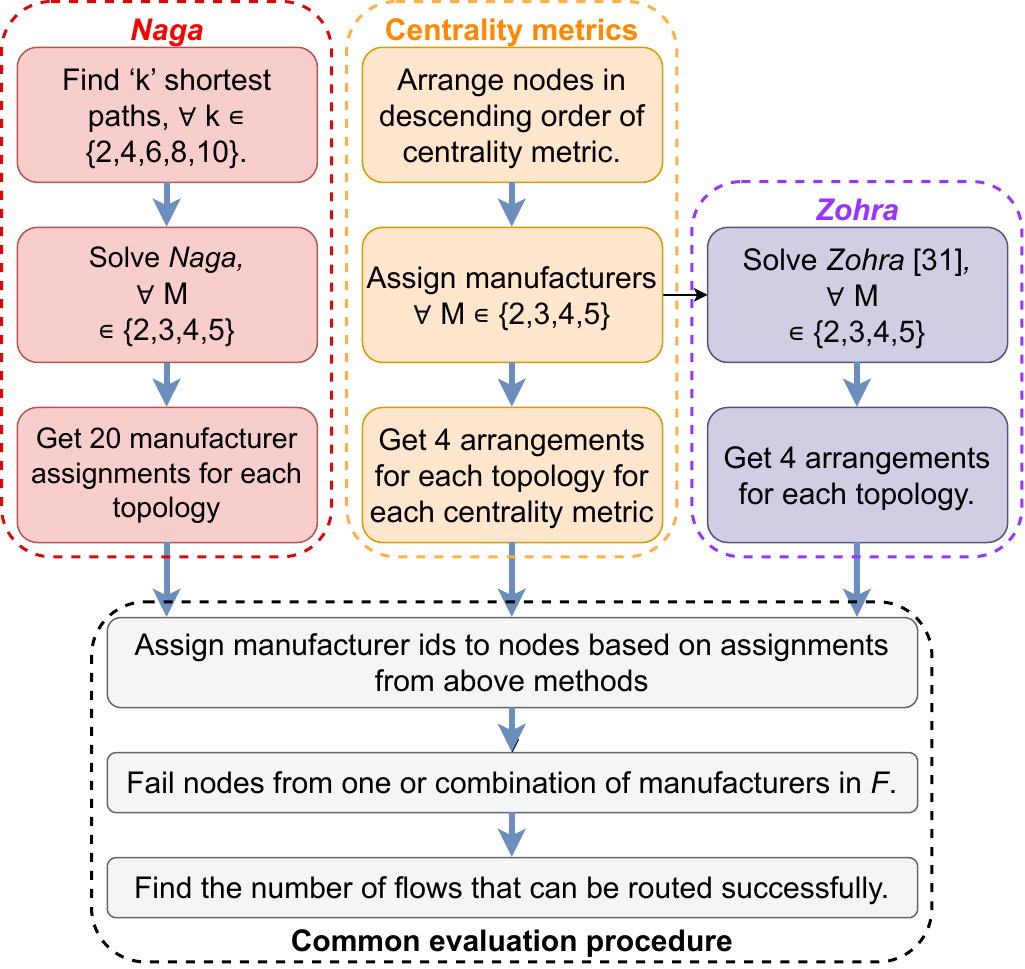}
\end{center}
\caption{Evaluation workflow of the proposed solution \textit{Naga} concerning the centrality metrics-based heuristics and the state of the art \textit{Zohra}.} 
\label{fig:process_Naga}
\end{figure}

\subsection{\textit{Naga} Evaluation}
\label{sec:nagaeval}
The procedure to evaluate \textit{Naga} is shown in the red box on the left in Fig.~\ref{fig:process_Naga}.
First, for \textit{Naga}, for each topology, for each flow, the $k$-shortest paths are found for $k \in \{2,4,6,8,10\}$. Then, for each $k$, four different number of manufacturers $|M| \in \{2,3,4,5\}$ are considered. Therefore, each topology has twenty different manufacturer assignments.

\subsection{Heuristics and \textit{Zohra} to compare with \textit{Naga}}
\label{sec:heuristics}
Since there are no works in the literature to quantify or maximize network sovereignty, we consider manufacturer assignments based on heuristics for comparison. The assignments could be done according to (i) centrality metrics (described in Section~\ref{sec:centrality}) and (ii) availability (e.g., as proposed in \textit{Zohra}~\cite{mine4} and described in Section~\ref{sec:Zohra}). These alternatives will be used to compare \textit{Naga}'s performance.

\subsubsection{Centrality metrics-based heuristics}
\label{sec:centrality}
Centrality metrics measure how important a node is in the network. If all the important nodes in the network are purchased from the same manufacturer and if that manufacturer is unavailable, this could be catastrophic for the network because several flows would fail. Hence, an operator would not want to assign nodes of high importance to the same manufacturer to avoid a dependency on that manufacturer. Therefore, we assign nodes of similar importance to different manufacturers to avoid dependencies.  

The evaluation process flow for the centrality metrics is shown in 
the center in Fig.~\ref{fig:process_Naga}. 
The centrality metric value for each node is calculated. Then, the nodes are arranged according to their centrality metric values. Then, the nodes with similar centrality metric values are assigned alternatively to different manufacturers.

For example, consider the graph $G^\ast$ in Fig.~\ref{fig:psdexample} with nodes $\{S,A,B,C,D,E,F,T\}$. Consider the nodal degree (ND) centrality metric for the manufacturer assignment in this example. The nodal degrees of the nodes are $\{\text{{ND}}_S = 4, \text{ND}_A = 3, \text{ND}_B = 3, \text{ND}_C = 2, \text{ND}_D = 2, \text{ND}_E = 3, \text{ND}_F = 2, \text{ND}_T = 3\}$. 
This list is rewritten in descending order of ND as $\{S,A,B,E,T,C,D,F\}$.
If we consider two manufacturers, $|M|=2$, the nodes bought from them are divided into $M_0 = \{S,B,T,D\}$ and $M_1 = \{A,E,C,F\}$. Similarly, if $|M| = 3$, then, $M_0 = \{S, E, D\}$, $M_1 = \{A,T,F\}$, and $M_2 = \{B,C\}$.
In such an assignment, nodes of similar importance are distributed among different manufacturers, making the network comparatively more sovereign than grouping all the important nodes into the same manufacturer.

While using centrality metrics, $k$ plays no role as it does not affect the centrality metric calculations. Therefore, it only depends on the number of manufacturers $|M| \in \{2,3,4,5\}$. To make this comparison more comprehensive, three different centrality metrics~\cite{saxena2020centrality} are considered- Nodal Degree (ND), Betweenness Centrality (BwC), and Closeness Centrality (CC). The BwC is calculated for each node only with respect to the source-destination pairs that have traffic present in the network. However, in our case study, any-to-any traffic is considered as stated in Section~\ref{sec:inputpara}. The arrangements for BwC and CC follow the same procedure as with ND. 

\subsubsection{\textit{Zohra}- Manufacturer assignment for availability}
\label{sec:Zohra}
The evaluation process flow for the manufacturer assignment from \textit{Zohra}~\cite{mine4} is shown in the purple box on the right in Fig.~\ref{fig:process_Naga}. The availabilities of the different manufacturers in this work are as per the considerations in \cite{mine4}. \textit{Zohra} is run for different topologies for different $|M| \in \{2,3,4,5\}$. Therefore, each topology has four different manufacturer assignments from \textit{Zohra}. Note that \textit{Zohra} was developed to maximize network availability and not sovereignty. It is evaluated in this work to compare the assignment for sovereignty against the one for availability.

\subsection{Evaluation Procedure}
\label{sec:evalproc}
After the manufacturer assignments are obtained, all possible combinations of manufacturer failures are simulated to evaluate the number of flows that can be successfully routed. For example, when there are three manufacturers, $M = \{0, 1, 2\}$, the possible failure combinations $F$ are $\{[0], [1], [2], [0,1], [0,2], [1,2]\}$. It is not meaningful to consider all manufacturers failing simultaneously.
When one manufacturer is unavailable, all nodes associated with that manufacturer are unavailable. Then, the percentage of successful traffic in each case is measured. A fully sovereign network must be able to route all the traffic if at least one manufacturer is functional. However, this is impossible to achieve in a real-world network. Therefore, a sovereign network must provide a working path for the maximum number of flows. This process is similar to the homogeneous analysis discussed in Section IV-4 in~\cite{mine1}. The common procedure to evaluate the different assignments from \textit{Naga}, centrality metrics, and \textit{Zohra} is shown in the lower black box in Fig.~\ref{fig:process_Naga}.

Note that source or destination failures are not considered because this is not an availability study. Additionally, we do not measure the network's performance degradation due to the failures. 
In a sovereignty study, the focus is only on the connectivity of the source-destination pairs.

%% file: 6_Results.tex
\section{Results}
\label{chap:results}
This section discusses our results. First, we identify some patterns in \textit{Naga}'s manufacturer assignment. Then, we discuss the PSD score and how \textit{Naga} improves it. Additionally, we provide guidelines on the optimal values of the number of manufacturers $|M|$ and the number of shortest paths $k$.

\subsection{Patterns in \textit{Naga}'s manufacturer assignment}
\begin{figure*}[tb]
\begin{subfigure}{0.24\linewidth}
    \includegraphics[width=1\linewidth]{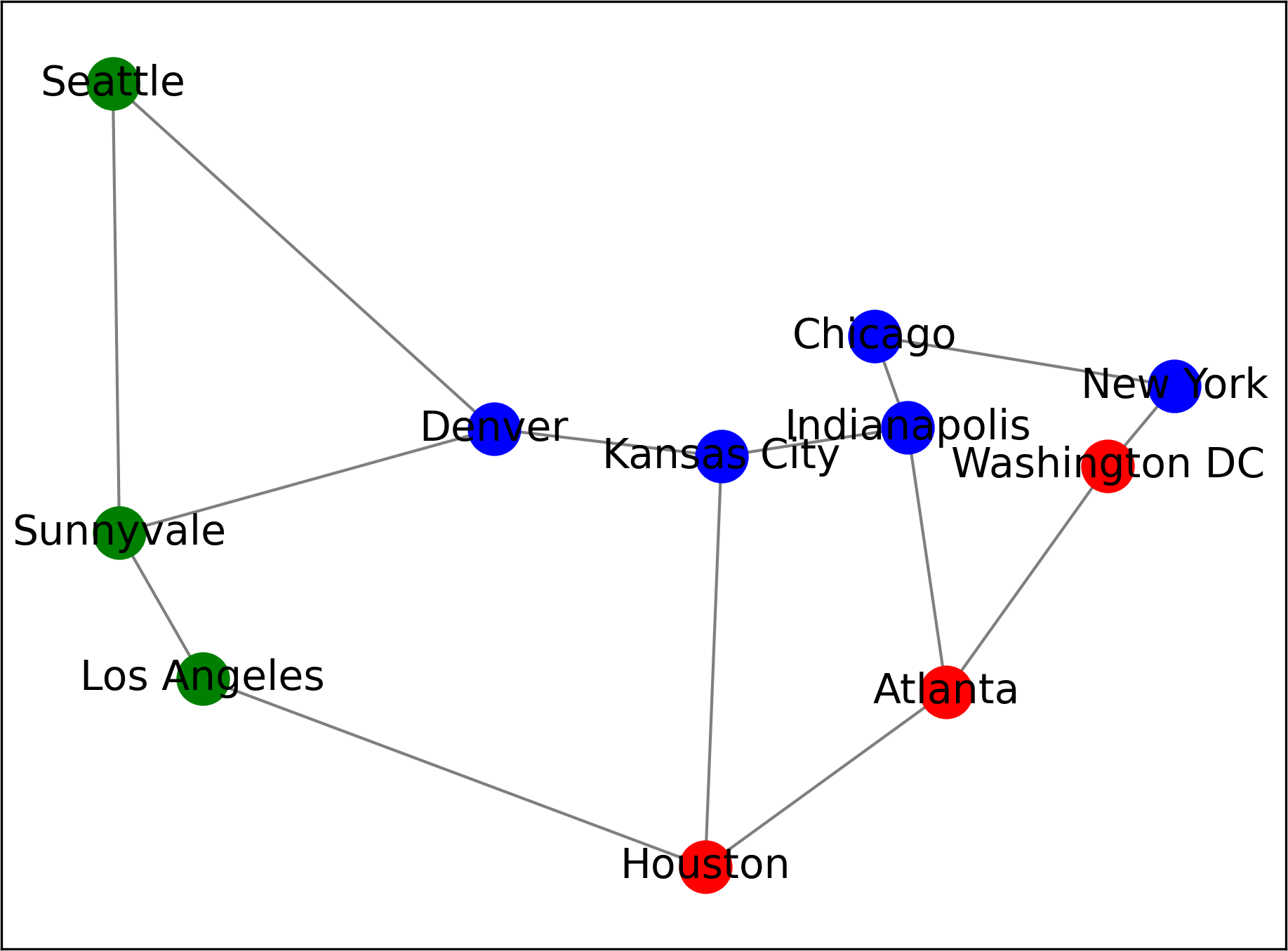}
    \caption{\textit{Naga} for $k=8$.} 
    \label{fig:color1}
\end{subfigure}
\begin{subfigure}{0.24\linewidth}
    \includegraphics[width=1\linewidth]{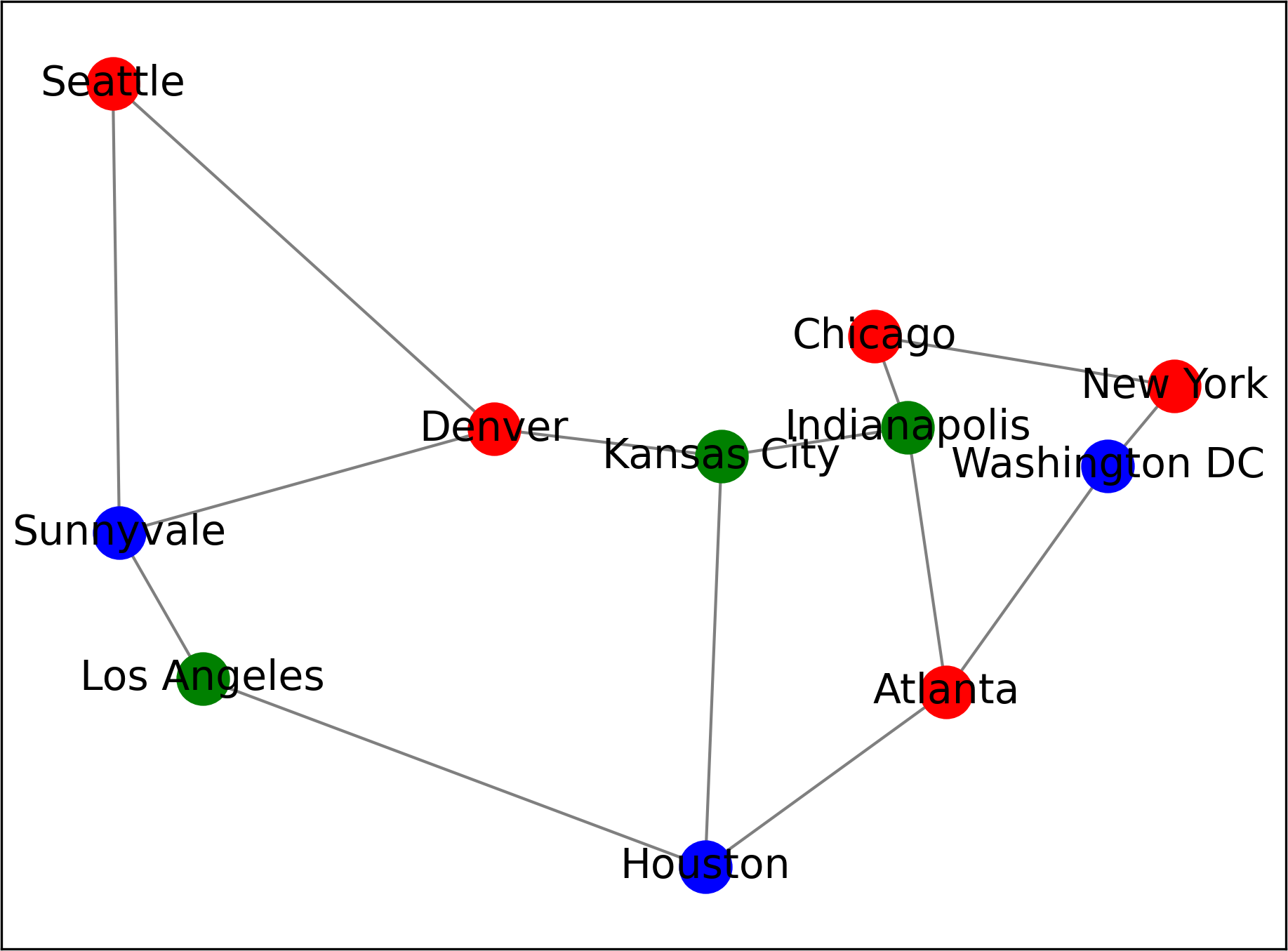}
    \caption{Nodal degree.}  
    \label{fig:color2}
\end{subfigure}
\begin{subfigure}{0.24\linewidth}
    \includegraphics[width=1\linewidth]{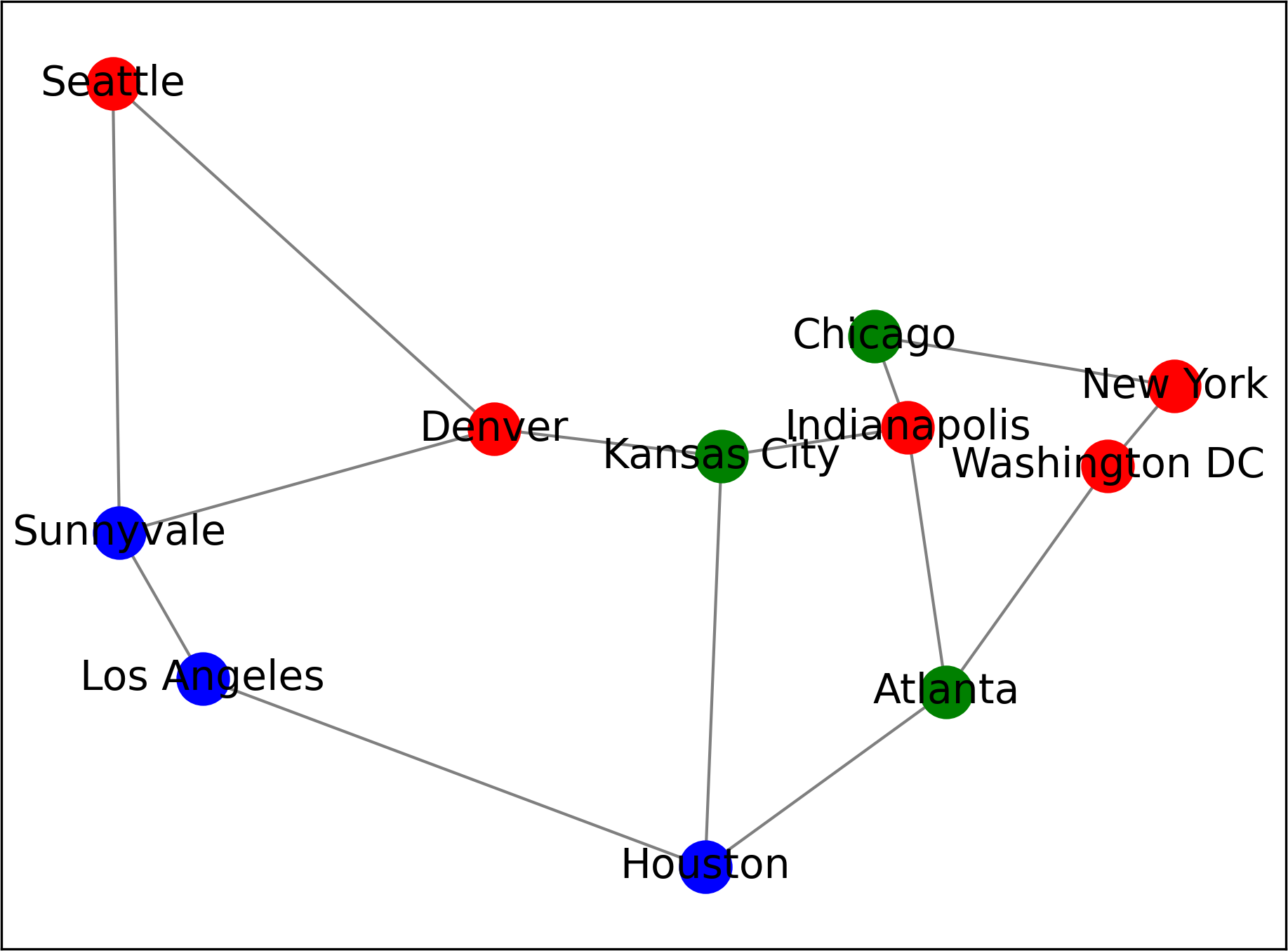}
    \caption{Betweenness centrality.}
    \label{fig:color3}
\end{subfigure}
\begin{subfigure}{0.24\linewidth} 
    \includegraphics[width=1\linewidth]{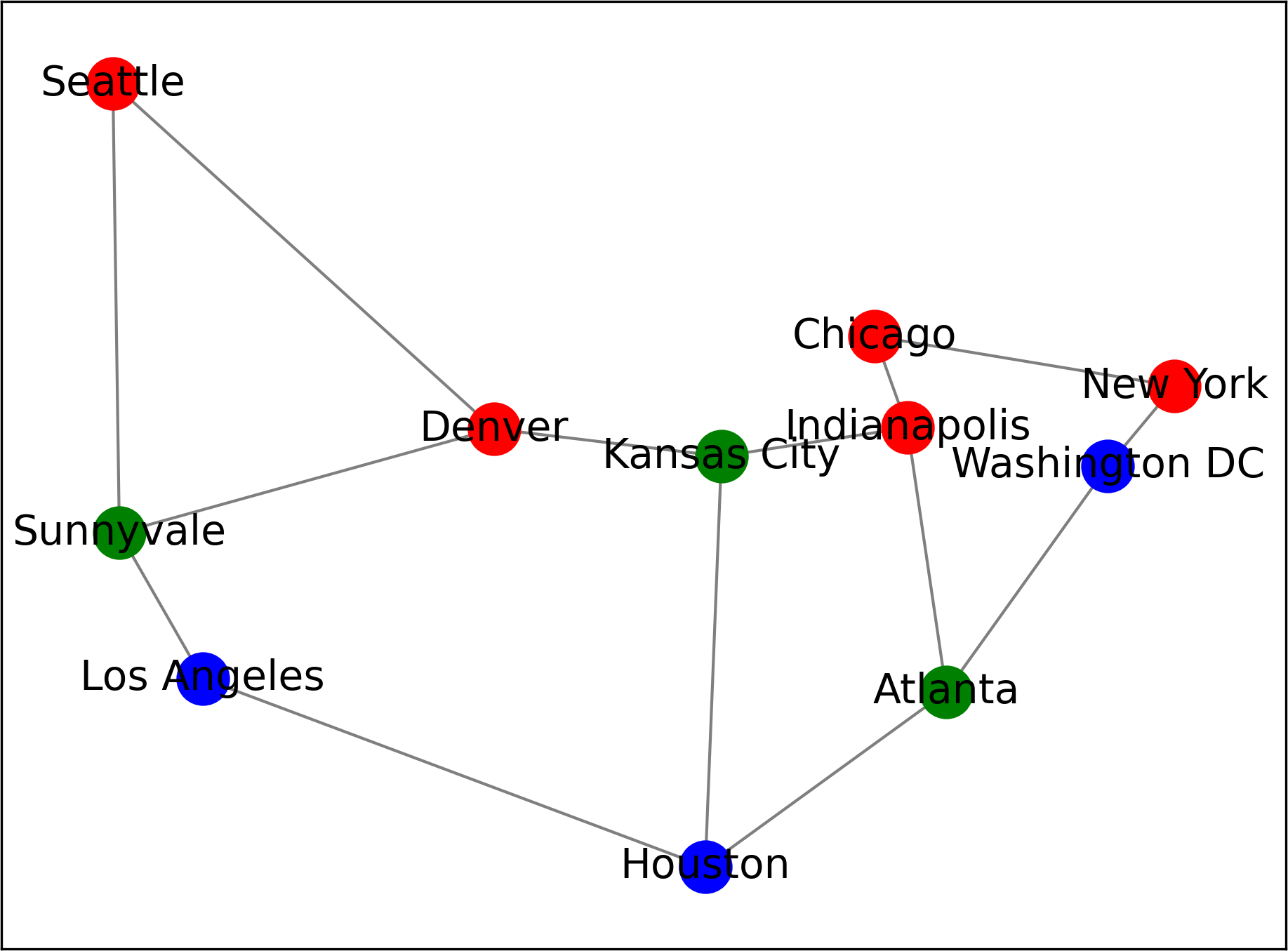}
    \caption{Closeness centrality.}
    \label{fig:color4}
\end{subfigure}
\caption{Manufacturer assignment on Abilene, $|M| = 3$.}
\label{fig:colors}
\end{figure*}

In Fig.~\ref{fig:colors}, we show the different manufacturer assignments as color coding in the Abilene topology with three manufacturers as an example. Fig.~\ref{fig:color1} shows the manufacturer assignment based on \textit{Naga}, for eight shortest paths ($k = 8$), Figs.~\ref{fig:color2},~\ref{fig:color3}, and~\ref{fig:color4} show the assignments based on ND, BwC, and CC respectively. Interestingly, there appears to be a chain-like pattern seen only in the assignment from \textit{Naga}. For example, in Fig.~\ref{fig:color1}, Seattle, Sunnyvale, and Los Angeles form a chain; Denver, Kansas City, Indianapolis, Chicago, and New York form a second chain; and Washington D.C., Atlanta, and Houston form a third chain. Such chain-like contiguous manufacturer assignment patterns are seen throughout the different topologies for different $|M|$ and $k$ values. 

This chain-like pattern ensures that each node is immediately connected to as many nodes from different manufacturers as possible. Therefore, even if one manufacturer fails, the connected nodes from other manufacturers still work. 
For example, in Fig.~\ref{fig:color1}, Houston is connected to a red node, a blue node, and a green node. 
However, such a chain-like pattern is not observed in the centrality metrics-based assignments. 
For example, in Figs.~\ref{fig:color2},~\ref{fig:color3}, and~\ref{fig:color4}, Houston is connected to at least two nodes from the green manufacturer alone. This is because the ND, BwC, and CC-based arrangements do not try to maximize the path diversity.

\subsection{Discussion on the PSD score and \textit{Naga}}
\label{sec:valid}

The goal of this section is to show that,
\begin{enumerate}[label=({V}\arabic*),align=left]
	\item\label{statement:V1} PSD score is an indicative metric of network sovereignty,
        \item\label{statement:V3} The improvement in sovereignty provided by the PSD score-based manufacturer assignment- \textit{Naga} is substantial compared to the centrality metrics-based heuristics and \textit{Zohra}.
\end{enumerate}
Both of these statements are interconnected; one can not be explained without the other.

\begin{figure}[tb]
    \begin{center}
    \includegraphics[width=0.9\linewidth]{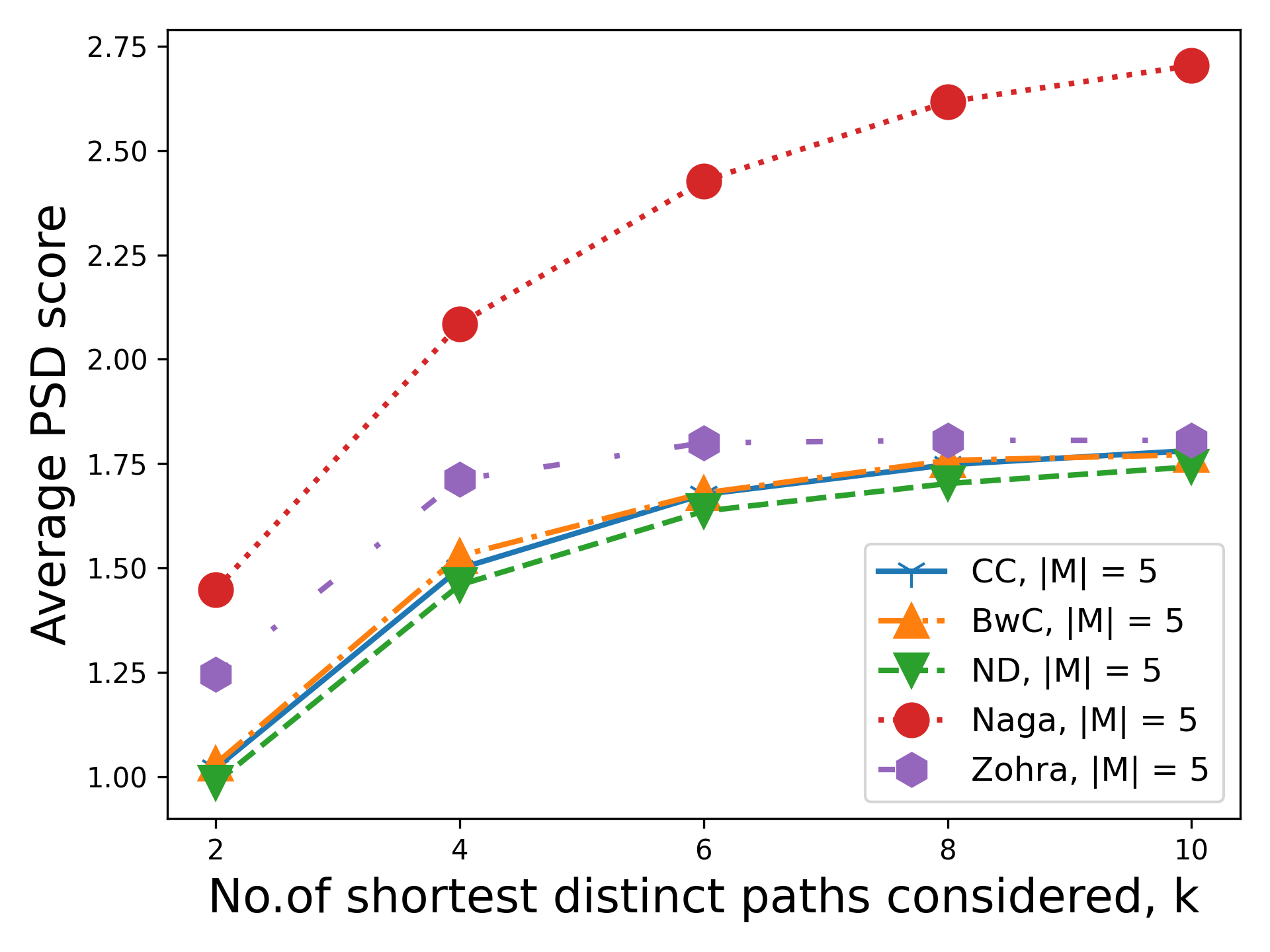}
    \end{center}
    \caption{PSD score vs. Number of shortest paths ($k$), Abilene, $|M| = 5$- \textit{Naga} has higher PSD scores than centrality metrics and \textit{Zohra}.} 
    \label{fig:r1}
\end{figure}

First, to show that the PSD score is an indicative metric of network sovereignty, we must show that the percentage of successful flows under failure scenarios is the highest for the PSD score-based manufacturer assignment from \textit{Naga}. To test that \textit{Naga} maximizes the PSD score, we compare the PSD score of the manufacturer assignment from \textit{Naga} with the PSD score of the manufacturer assignments obtained from the centrality metrics and \textit{Zohra}.

Fig.~\ref{fig:r1} shows the PSD scores of manufacturer assignments from \textit{Naga}, the three centrality metrics, and \textit{Zohra} for the Abilene topology. The Y-axis shows the PSD score. The X-axis shows the possible number of shortest paths, $k$. Fig.~\ref{fig:r1} shows that \textit{Naga}'s manufacturer assignment has the best PSD score irrespective of the number of shortest paths considered.

\begin{figure}[tb]
\begin{subfigure}{1\linewidth}
    \begin{center}
    \includegraphics[width=0.9\linewidth]{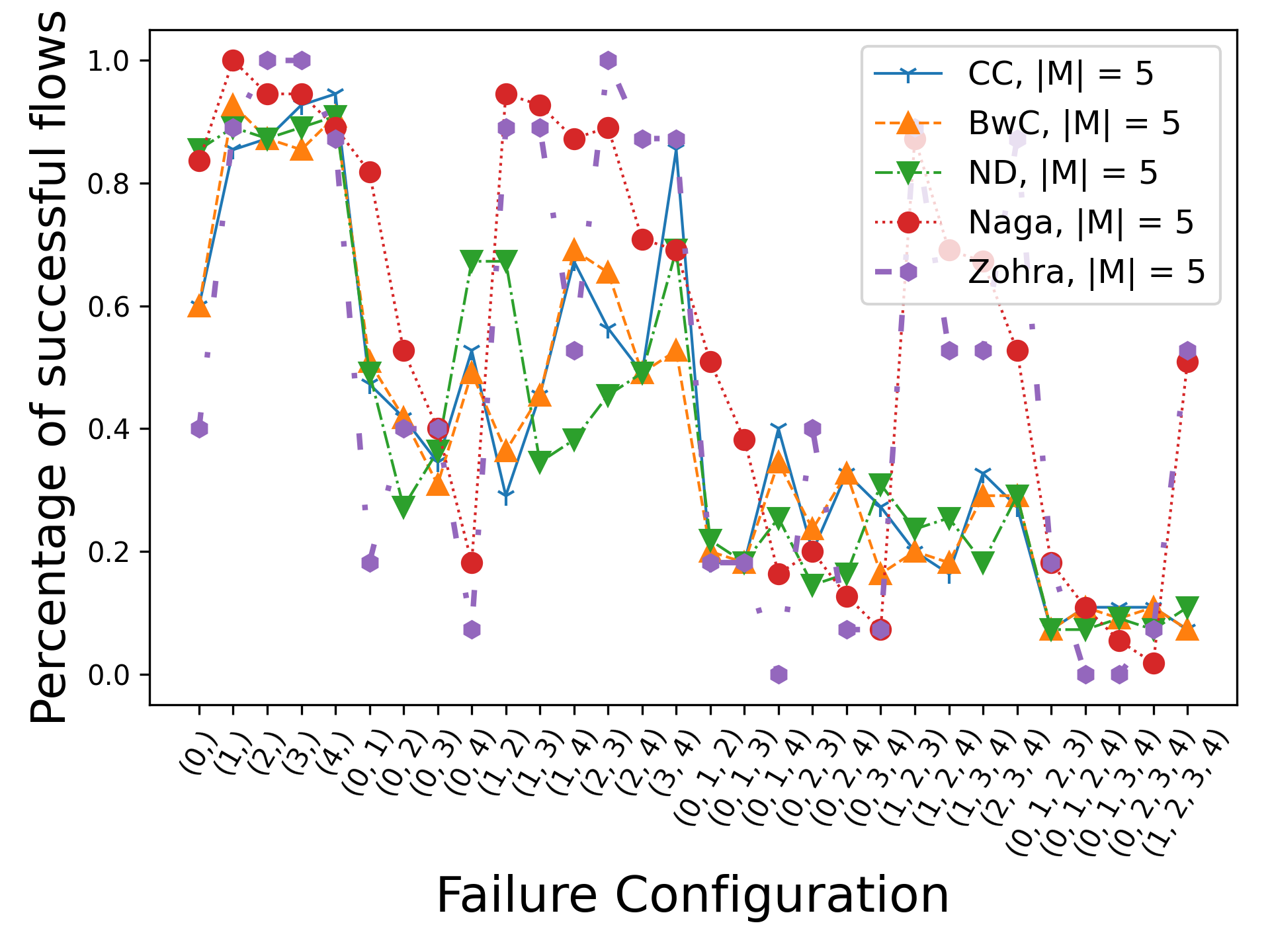}
    \end{center}
    \caption{Percentage of successful flows vs. Failure scenarios.}  
    \label{fig:rb1}
\end{subfigure}
\begin{subfigure}{1\linewidth}
    \begin{center}
    \includegraphics[width=0.9\linewidth]{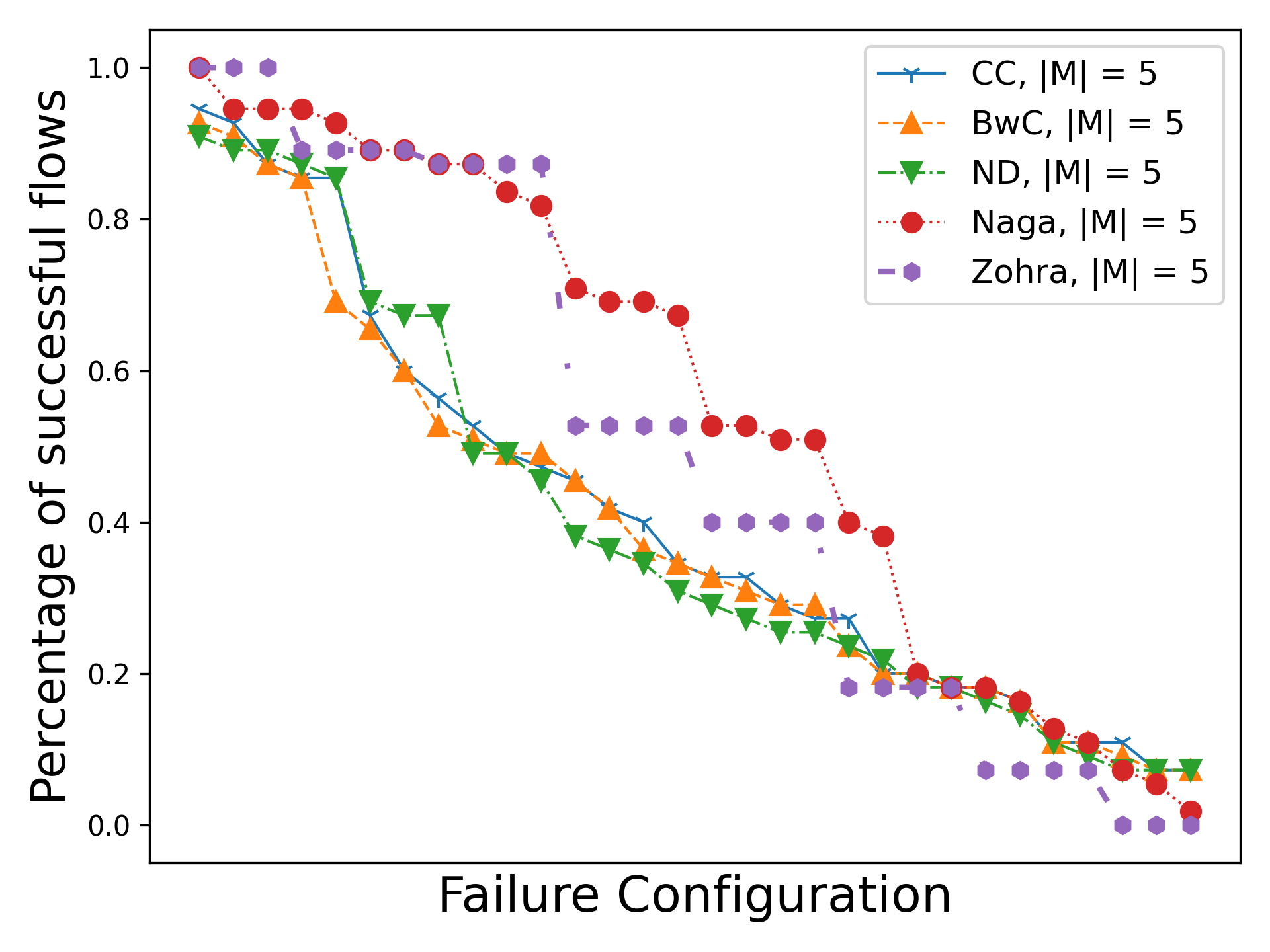}
    \end{center}
    \caption{Percentage of successful flows in descending order.}  
    \label{fig:r2}
\end{subfigure}
\caption{\textit{Naga}'s performance: Abilene, $|M| = 5$, $k = 10$- \textit{Naga} has more successful flows under failure scenarios than centrality metrics and \textit{Zohra}.}
\label{fig:example1}
\end{figure}

Fig.~\ref{fig:rb1} shows the percentage of successful flows under different manufacturer failure scenarios for the Abilene topology for five manufacturers ($|M| = 5$) and ten shortest paths considered ($k = 10$). The Y-axis shows the percentage of successful flows, while the failure scenarios are on the X-axis. This graph shows that \textit{Naga} has more successful flows than the other assignments under failure scenarios. However, to clearly show \textit{Naga}'s performance, we plot Fig.~\ref{fig:r2} based on Fig.~\ref{fig:rb1}. In Fig.~\ref{fig:r2}, the different percentages of successful flows are obtained for each manufacturer assignment like Fig.~\ref{fig:rb1}, and then, the points in each line are arranged in descending order. Therefore, it only shows an assignment's relative advantage (or disadvantage) and does not give direct failure scenario comparisons like shown in Fig.~\ref{fig:rb1}. Since the lines are all individually arranged in descending order, their X-axes are no longer the same. Hence, there are no X-axis ticks in Fig.~\ref{fig:r2}.
Fig.~\ref{fig:r2} specifically shows that \textit{Naga} has more successful flows than the other assignments under failure scenarios. Therefore, \textit{Naga}'s assignment is the best for sovereignty. A step-like pattern is observed for \textit{Zohra} because \textit{Zohra}, an availability optimization, may produce a manufacturer assignment that does not use a particular manufacturer of the five available manufacturers. Therefore, if a manufacturer not used in \textit{Zohra}'s assignment fails, the Y-axis has no impact.

\begin{figure}[tb]
\begin{subfigure}{1\linewidth}
    \begin{center}
    \includegraphics[width=0.9\linewidth]{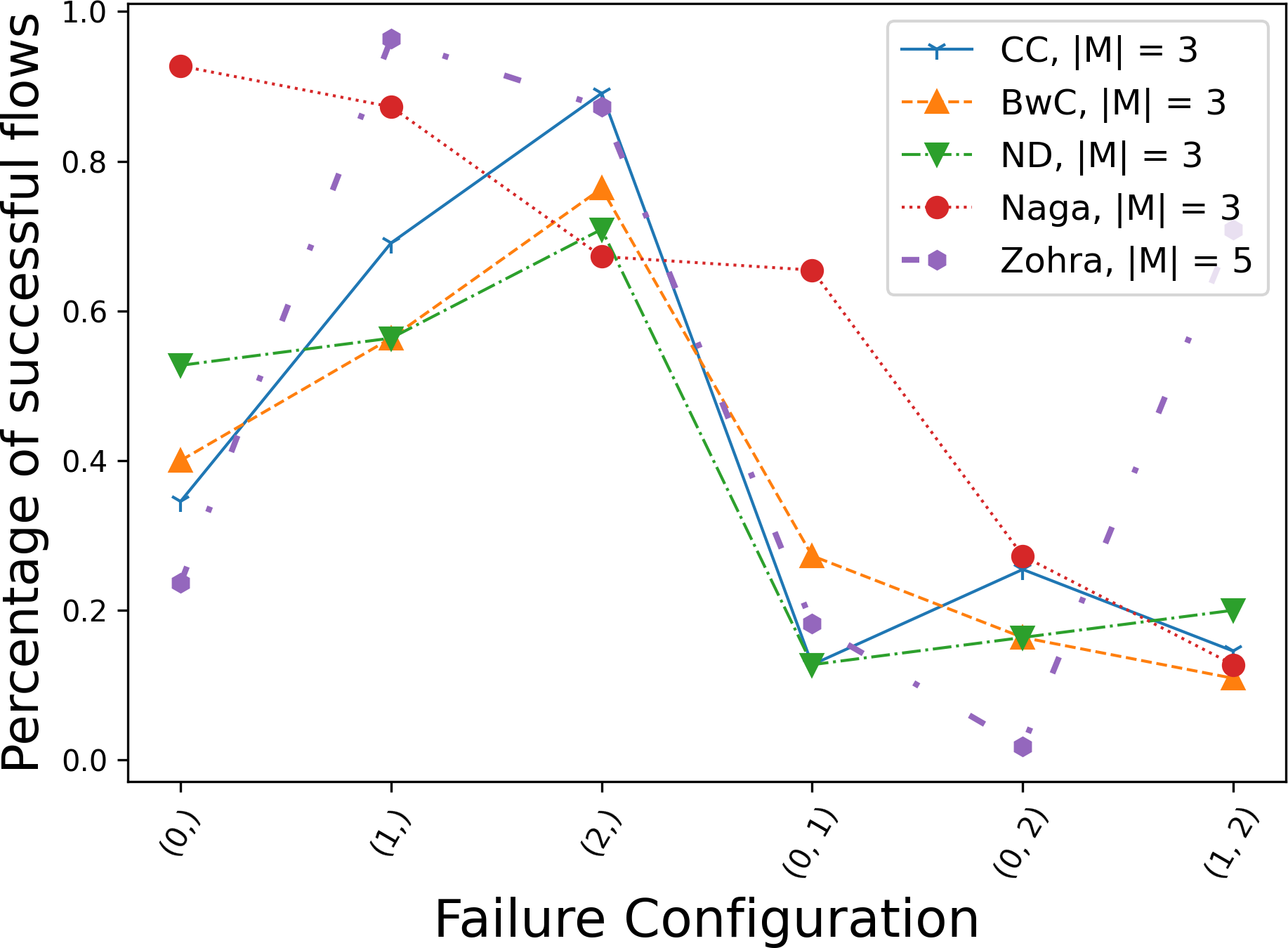}
    \end{center}
    \caption{Percentage of successful flows vs. Failure scenarios.} 
    \label{fig:rb2}
\end{subfigure}
\begin{subfigure}{1\linewidth}
    \begin{center}
    \includegraphics[width=0.9\linewidth]{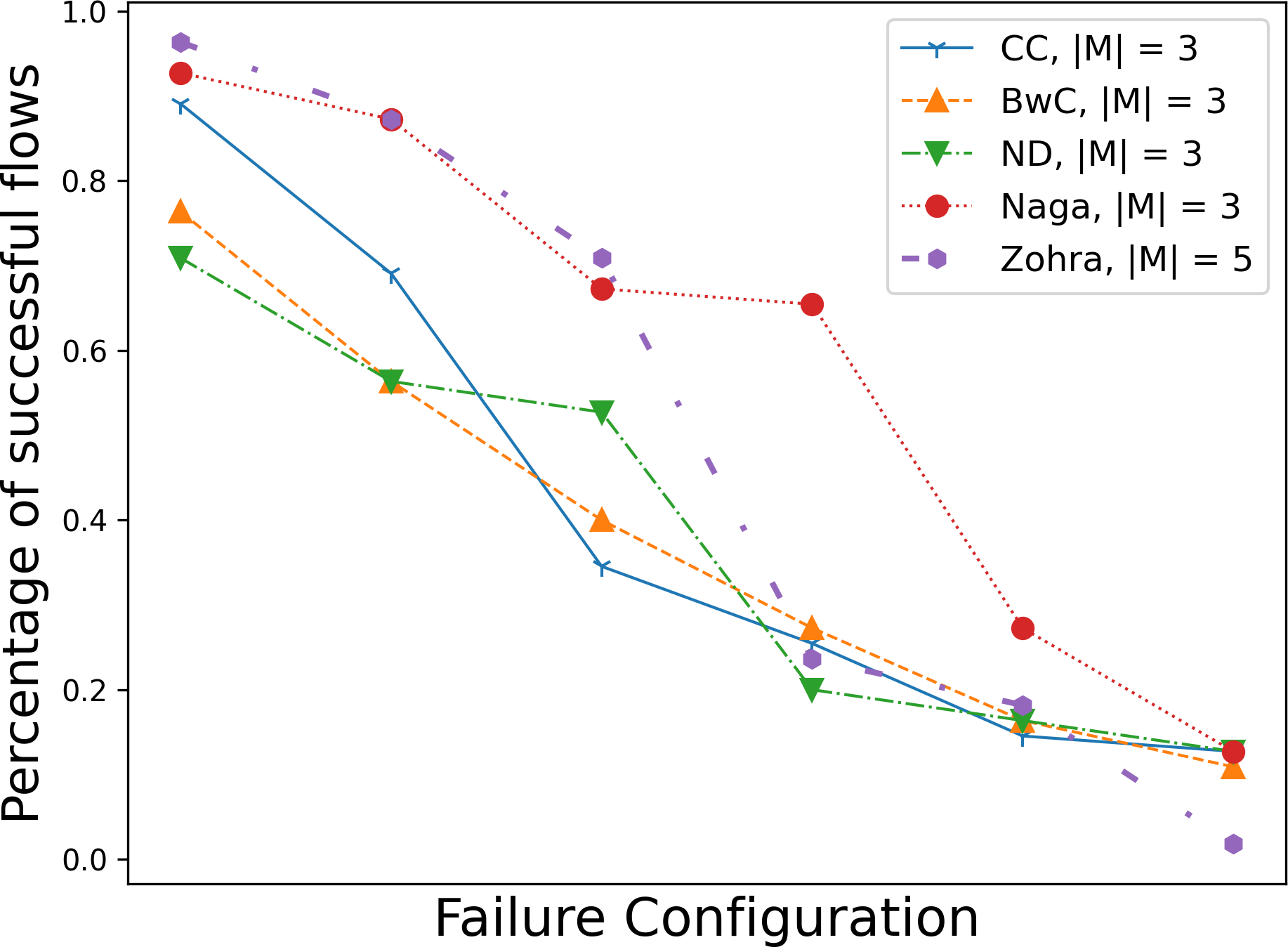}
    \end{center}
    \caption{Percentage of successful flows in descending order.}  
    \label{fig:r12plus}
\end{subfigure}
\caption{\textit{Naga}'s performance: Abilene, $|M| = 3$, $k = 6$- \textit{Naga} has more successful flows under failure scenarios than centrality metrics and \textit{Zohra}.}
\label{fig:example2}
\end{figure}

Similar to Figs.~\ref{fig:rb1}~and~\ref{fig:r2}, Figs.~\ref{fig:rb2}~and~\ref{fig:r12plus} shows the percentage of successful flows under different manufacturer failure scenarios for the Abilene topology for three manufacturers ($|M| = 3$) and six shortest paths considered ($k = 6$). Once again, the percentage of successful flows is arranged in descending order for each manufacturer assignment in Fig.~\ref{fig:r12plus} to identify the advantage of \textit{Naga}. Similar to Fig.~\ref{fig:r2}, Fig.~\ref{fig:r12plus} also does not have any X-axis ticks. \textit{Naga} performs consistently better than the other manufacturer assignments for all topologies, $|M|$, and $k$ values.

In summary, we can observe that when the PSD score is high for an assignment, as seen in Fig.~\ref{fig:r1}, the number of successful flows under failure scenarios is also high, as seen in Figs.~\ref{fig:example1} and~\ref{fig:example2}. This positive correlation between the PSD score and the percentage of successful flows shows that the statements~\ref{statement:V1} and~\ref{statement:V3} are true. Therefore, the PSD score is a good indication of a network's sovereignty, and \textit{Naga} improves network sovereignty.

\subsection{Critical takeaways from \textit{Naga}}
\label{sec:take}



Figs.~\ref{fig:rb1}~and~\ref{fig:rb2} reveal crucial information about the manufacturer(s) who cause the most impactful failures.

For example, for Abilene with five manufacturers ($|M| = 5$) in Fig.~\ref{fig:rb1}, $M_0$ has the highest impact on failing, followed by $M_4$. Consequentially, any failure scenario with either of these two manufacturers failing results in higher lost traffic. The worst case possible is both $M_0$ and $M_4$ failing simultaneously, along with others. Even if only the two of them fail, more than $80\%$ of the traffic is lost. On the other hand, if $M_1$ and $M_2$ fail simultaneously, not even $10\%$ of the flows are failing. So, in this case, the operator must be careful about choosing $M_0$ and $M_4$. However, this labeling is subjective and may change when using a different input configuration.

For example, for Abilene with three manufacturers ($|M| = 3$) in Fig.~\ref{fig:rb2}, the failures of $M_0$ and $M_1$ do not affect the network to a large extent. This means that there is no dependency on these two manufacturers. On the other hand, when $M_2$ fails, nearly one-third of the traffic is lost. This shows that $M_2$ is critical to the network. Consequentially, any failure scenario that involves $M_2$ also loses a lot of traffic. This is vital information for an operator who wants to know the worst situations that must be avoided. Therefore, in this case, the operator must assign the most trustworthy manufacturer to the nodes in manufacturer $M_2$ when using \textit{Naga}. 

A crucial inference here is that no assignments can guarantee $100\%$ successful flows when one or more manufacturers are unavailable. The goal is to find an assignment where the traffic loss is minimal. However, the weight parameter $w_r$ in Eq.~\ref{eq:psd} and Eq.~\ref{eq:o1} can be modified to give higher priority to more critical flows such that they are more sovereign and survive multiple manufacturers' unavailability.

Another critical inference is that an availability-based manufacturer assignment like \textit{Zohra} is not sufficient to guarantee maximum sovereignty. The network operator needs to find the optimal compromise between the network availability and sovereignty that the network requires.

\subsection{Impact of number of manufacturers $|M|$ on sovereignty}
\label{sec:impactM}
\begin{figure}[ht!]
\begin{subfigure}{1\linewidth}
    \begin{center}
    \includegraphics[width=0.9\linewidth]{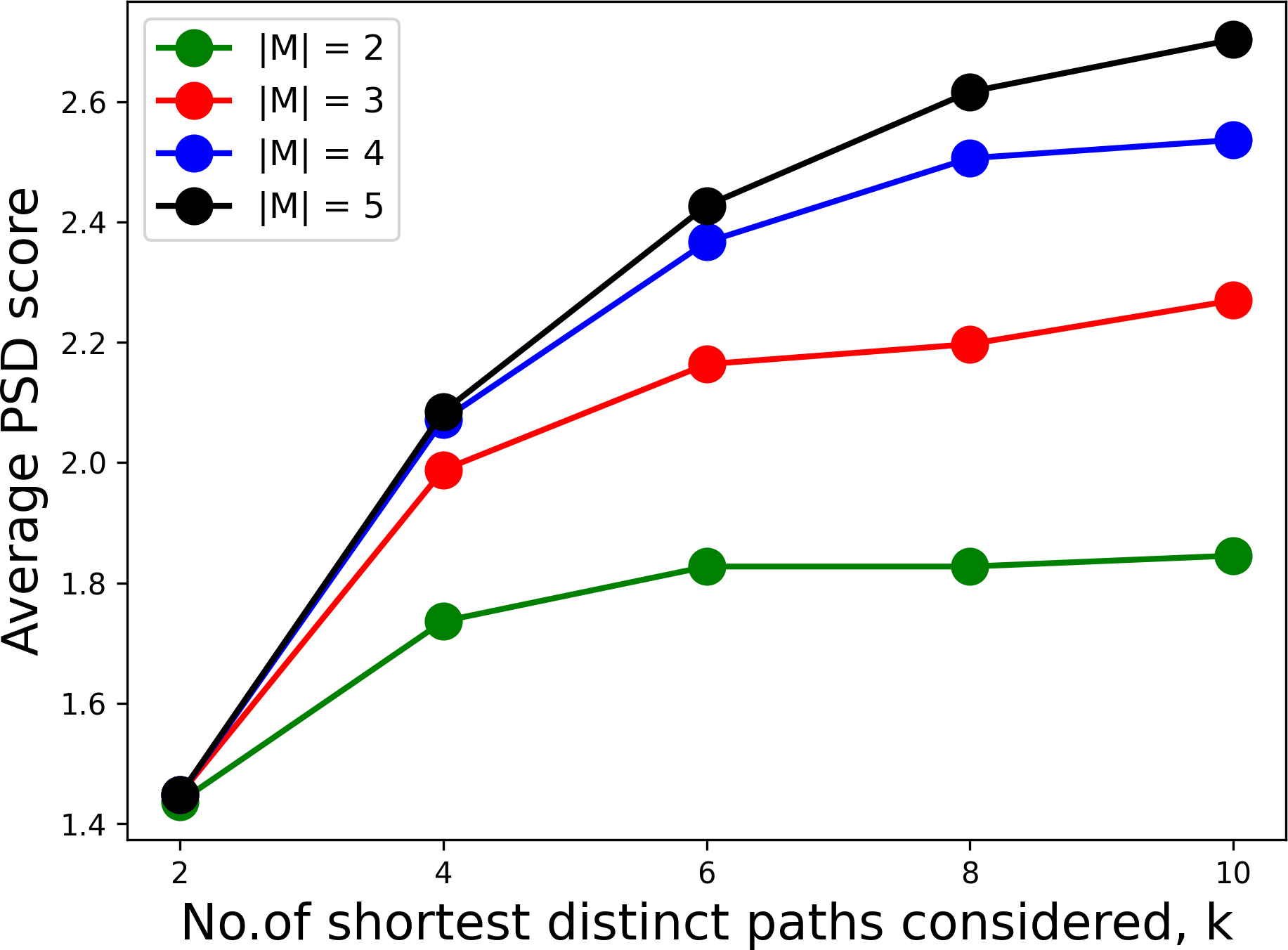}
    \end{center}
    \caption{PSD score comparison for different $|M|$, Abilene.} 
    \label{fig:mcomp1}
\end{subfigure}
\begin{subfigure}{1\linewidth}
\begin{center}
\includegraphics[width=0.9\linewidth]{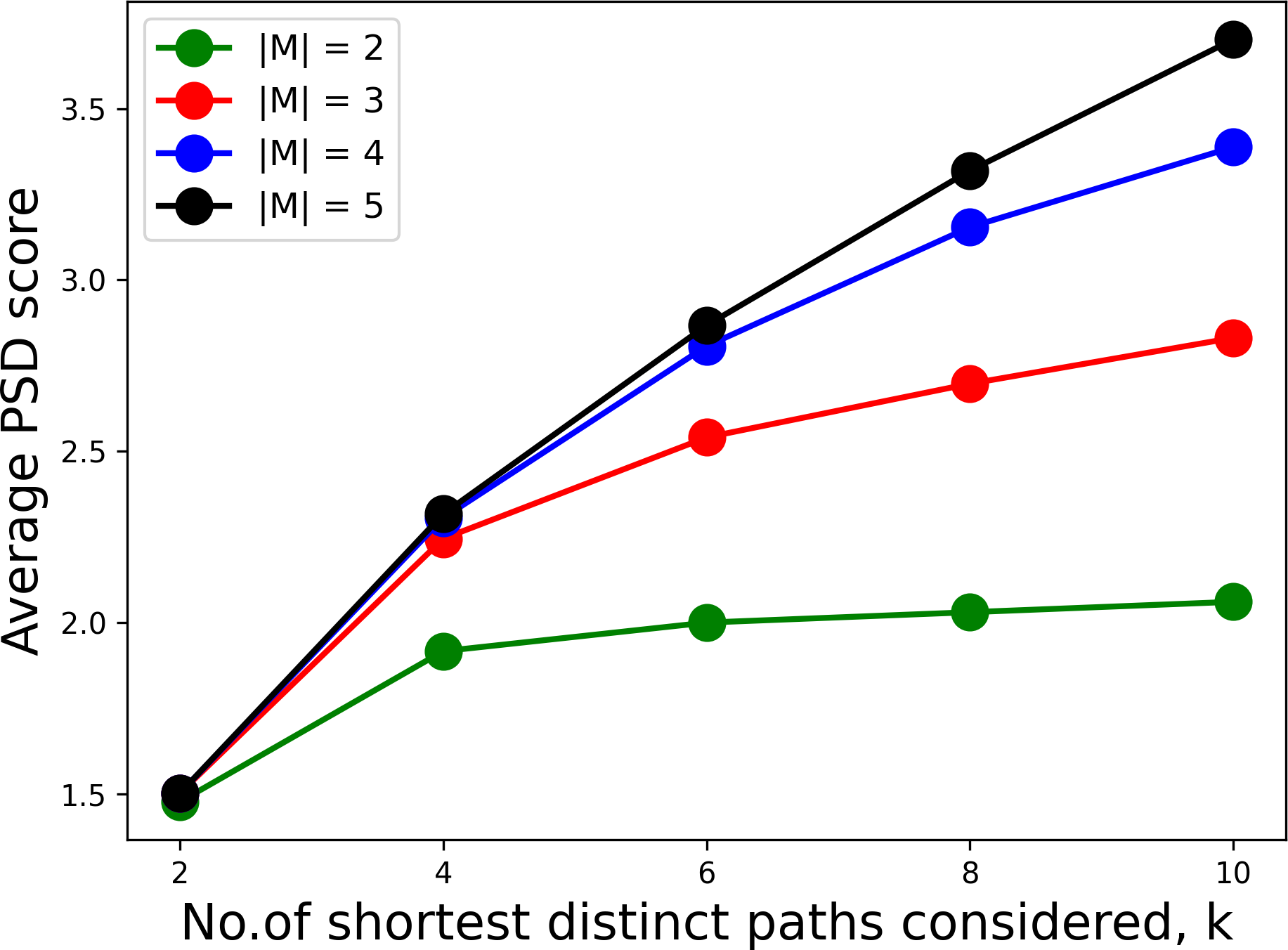}
\end{center}
\caption{PSD score comparison for different $|M|$, Polska.} 
\label{fig:mcomp-polska}
\end{subfigure}
\begin{subfigure}{1\linewidth}
    \begin{center}
    \includegraphics[width=0.9\linewidth]{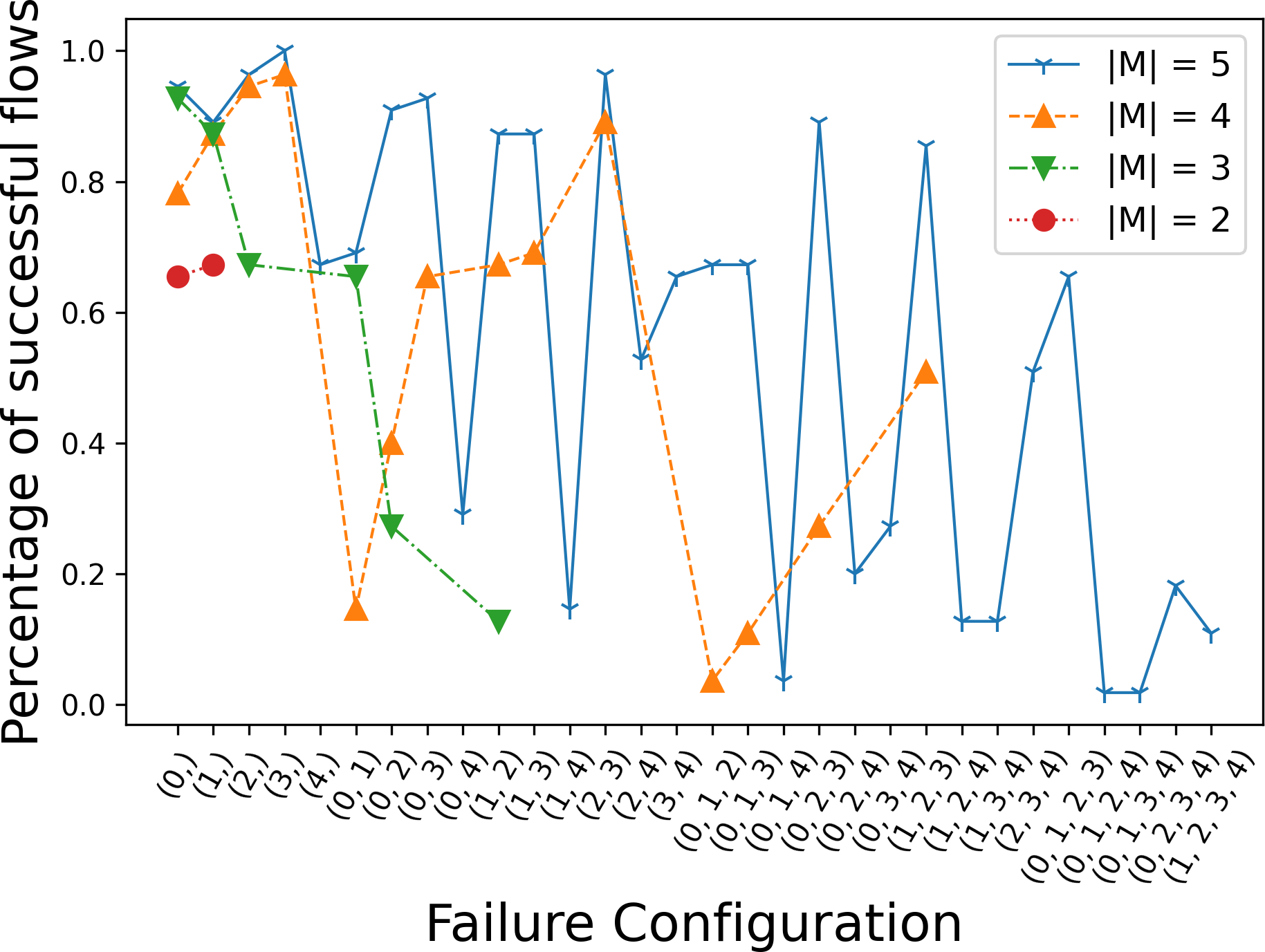}
    \end{center}
    \caption{Percentage of successful flows comparison for different $|M|$, $k = 6$, Abilene.}  
    \label{fig:mcomp2}
\end{subfigure}
\caption{Impact of number of manufacturers $|M|$- Higher $|M|$ corresponds to higher PSD score.}
\label{fig:mcomp}
\end{figure}

Now that we have established the performance of the PSD score and \textit{Naga}, we can continue with the network sovereignty analysis. First, we examine the impact of using more (or lesser) manufacturers in the network with Fig.~\ref{fig:mcomp}. 

Fig.~\ref{fig:mcomp1} shows the PSD score comparison for different numbers of manufacturers $|M|$ for the Abilene topology. The Y-axis and X-axis have the PSD scores and the number of shortest paths $k$, respectively. As expected, the PSD score is higher when $|M|$ is higher. This is because when $|M|$ is higher, there is less dependency on any single manufacturer. Furthermore, when $k$ increases for a given $|M|$, the PSD score also increases. This is also expected because when more paths are considered, more combinations of manufacturers in the paths will be considered for the PSD score, which should generally increase the PSD score. This increase slowly saturates at some point because, after a certain $k$, the same manufacturer combinations in paths are repeated, making them redundant and excluded from the PSD score calculation as per Line~6 in Algorithm~\ref{algo:psd}. Such characteristics are consistent across the topologies. For example, the same characteristics are also observed in Fig.~\ref{fig:mcomp-polska} for the Polska topology. However, due to more links in Polska, the saturation of the PSD score occurs later than in the Abilene topology.

Fig.~\ref{fig:mcomp2} shows the percentage of successful flows on the Y-axis and the different failure scenarios on the X-axis for different values of $|M|$. Only two failure scenarios are possible for $|M| = 2$, so it has only two points on the graph. As $|M|$ increases, more failure scenarios are possible. When more manufacturers are in the network, the network can tolerate more combinations of failures. Therefore, having more manufacturers and assigning them based on \textit{Naga} is essential to guarantee maximum network sovereignty.

\subsection{Impact of number of shortest paths $k$ on sovereignty}
\label{sec:impactK}
\begin{figure}[tb]
\begin{center}
\includegraphics[width=0.9\linewidth]{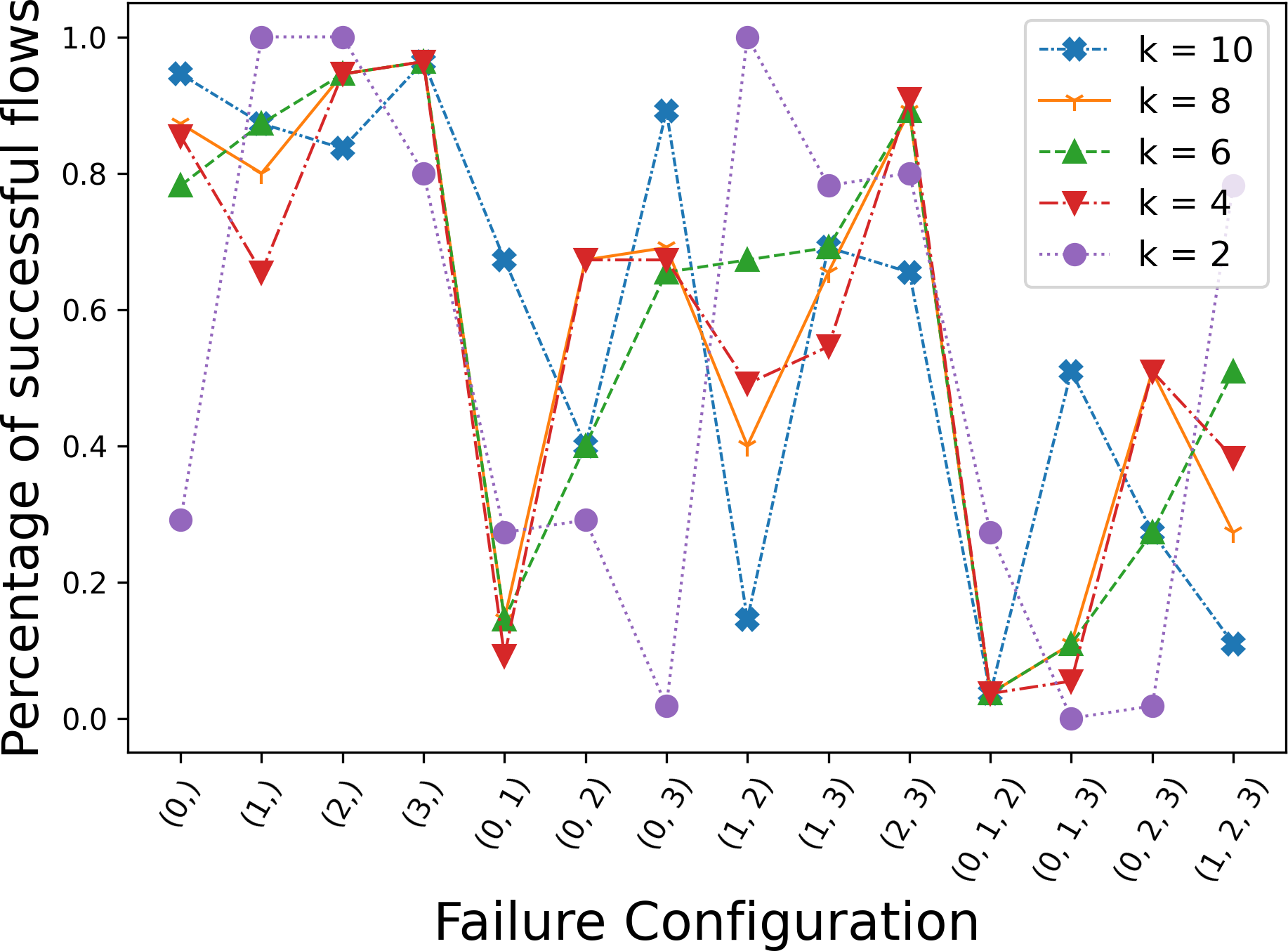}
\end{center}
\caption{Impact of number of shortest paths $k$: Abilene, $|M| = 4$- Lower $k$ corresponds to lesser successful flows under failure scenarios. However, this is difficult to quantify.} 
\label{fig:kcomp}
\end{figure}

In this section, we investigate the impact of choosing the number of shortest paths, $k$. Fig.~\ref{fig:kcomp} shows the $k$ comparison graph for the Abilene topology with four manufacturers ($|M| = 4$). The Y-axis has the percentage of successful flows. The X-axis has different failure configurations. Each line is for a particular value of $k$. Generally, having a lower $k$ has more failed flows than having a higher $k$. However, the difference in the percentage of successful flows is difficult to quantify for the different failure configurations and topologies. 

This graph also gives vital information about which failure scenarios cause the most severe failures for a particular value of $k$. This information can be used to avoid unreliable manufacturers in those positions. For example, when $k$ is two, $M_0$ causes the most failures. So, the most trustworthy manufacturer must be placed in this position. On the other hand, when $k$ is ten, for the failure of any single manufacturer, the worst performance is only a failure of less than 20\% of the flows. Here, $M_2$ gives the highest impact. 

Only results for some $|M|$ are graphically displayed as examples to handle space constraints. However, all the results in this section are consistent across different $|M|$ values.
From these results, we can confirm conclusively that we can answer the question~\ref{q:Q1} by giving the best manufacturer assignment possible to maximize network sovereignty.

\subsection{Special cases}
\label{sec:special}
\begin{figure}[tb]
\begin{subfigure}{1\linewidth}
\begin{center}
\includegraphics[width=0.9\linewidth]{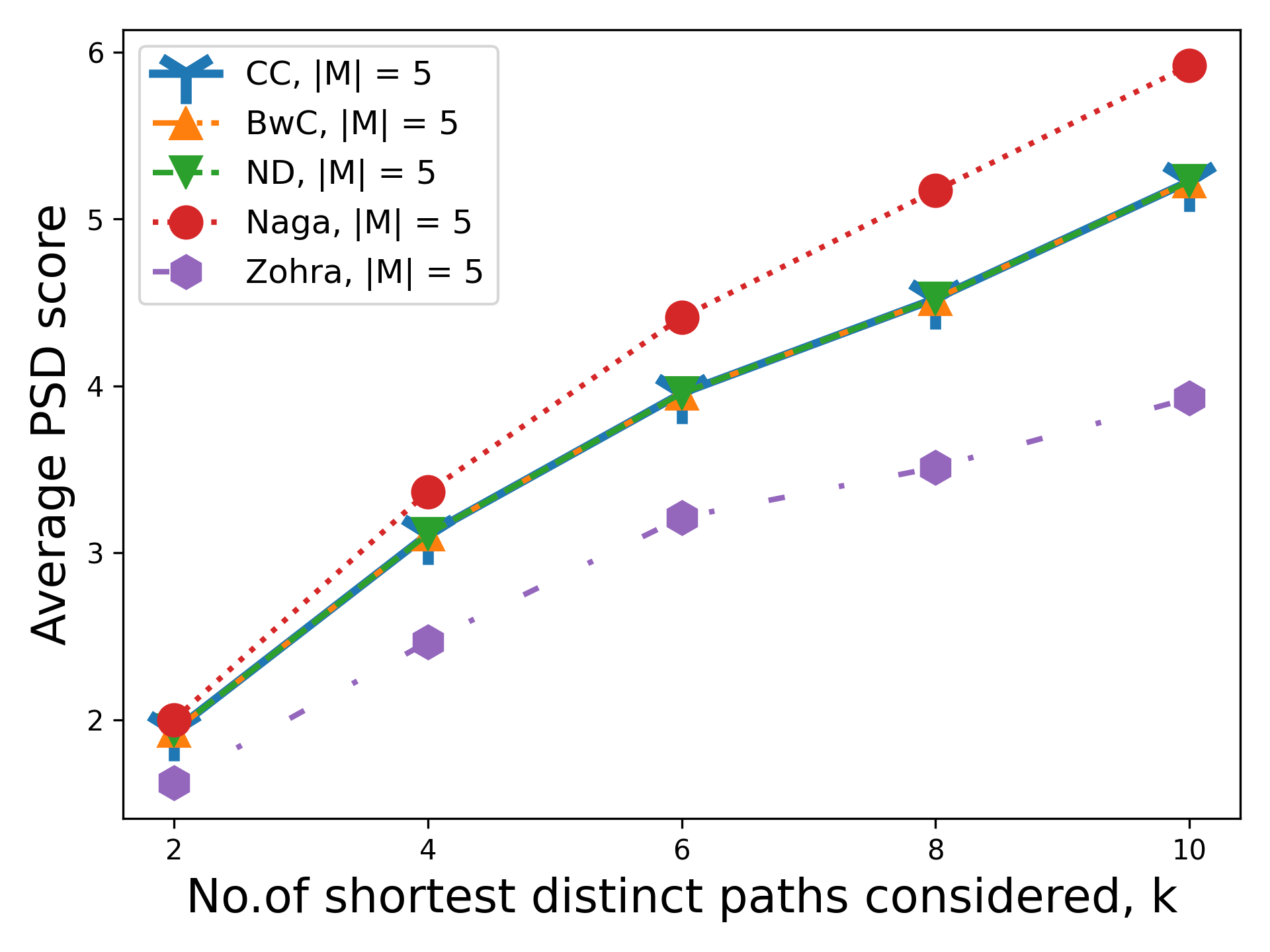}
\end{center}
\caption{PSD score vs. Number of shortest paths ($k$): dfn-bwin, $|M| = 5$- \textit{Naga} has the highest PSD score.} 
\label{fig:special1}
\end{subfigure}
\begin{subfigure}{1\linewidth}
\begin{center}
\includegraphics[width=0.9\linewidth]{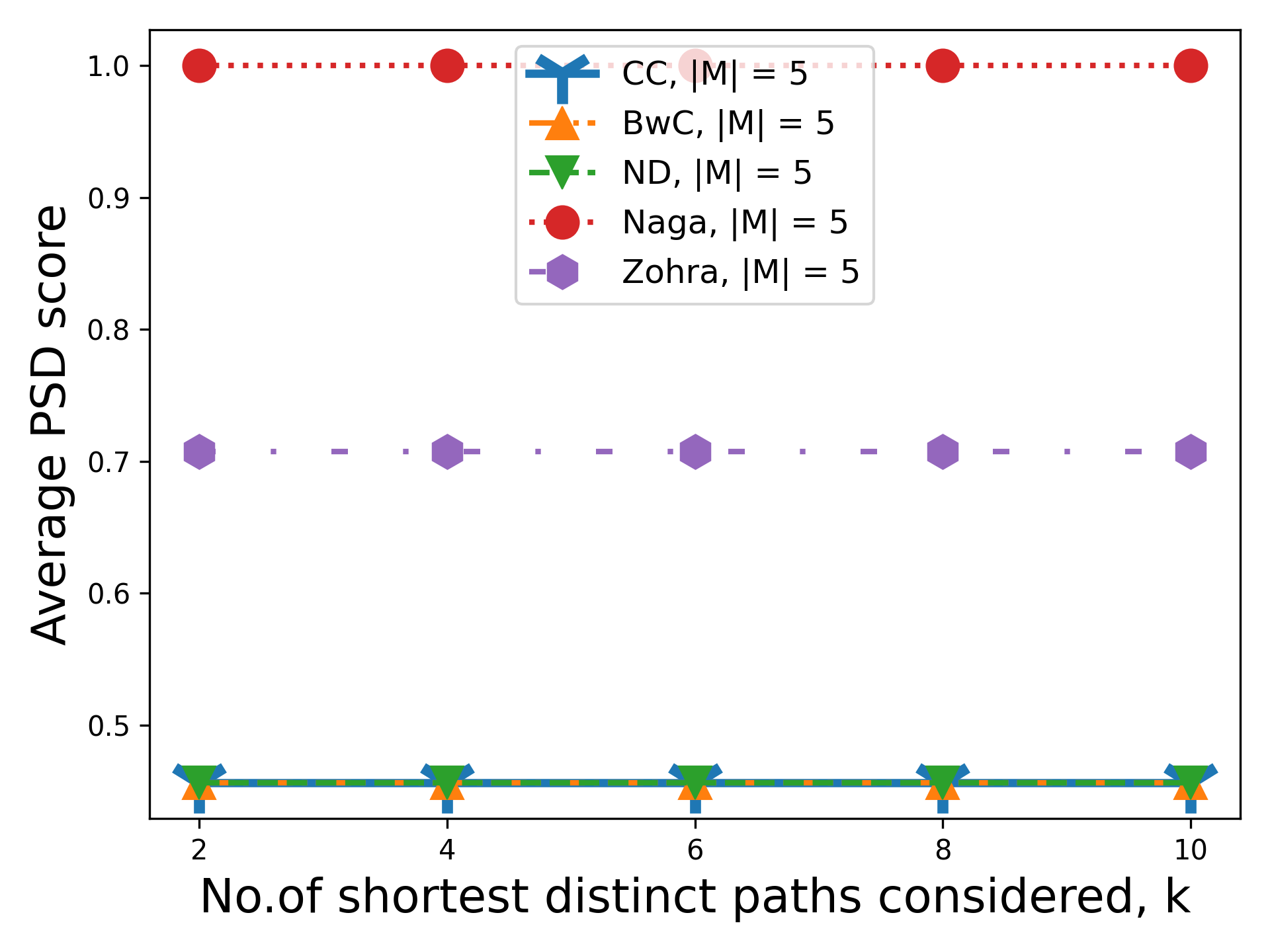}
\end{center}
\caption{PSD score vs. Number of shortest paths ($k$): HiberniaUK, $|M| = 5$- \textit{Naga} has the highest PSD score.} 
\label{fig:special2}
\end{subfigure}
\caption{Special cases- \textit{Naga} performs consistently even when centrality metrics are not valid.}
\label{fig:special}
\end{figure}
The Abilene and Polska topologies are conventional topologies, whereas dfn-bwin and HiberniaUK are special cases since they are mesh and ring topologies, respectively. In topologies like HiberniaUK with lower connectivity, the number of paths between a source and destination may be fewer than the $k$ value that is considered. In such a case, only the feasible paths for the source-destination pair are considered. This may lead to a lower $\pi_r$ agreeing with the guideline~\ref{statement:M1}, which states that the lack of multiple paths between source and destination must not be rewarded.

Fig.~\ref{fig:special1} shows the average PSD score compared with $k$, similar to Fig.~\ref{fig:r1}.
In dfn-bwin, the mesh topology, all the nodes have equal centrality metrics. Therefore, the ordering of nodes based on the centrality metrics mentioned in Section~\ref{sec:centrality} gives the same manufacturer assignment for all the centrality metrics. However, the manufacturer assignment due to \textit{Naga} varies because the number of shortest paths considered, $k$, influences \textit{Naga}'s solution. As a result, the PSD score also increases with an increase in $k$, consistent with all the previous results discussed in Section~\ref{chap:results}. 

However, HiberniaUK is a ring topology with only two paths between any source and destination. Therefore, $k$ does not influence the results at all. Fig.~\ref{fig:special2} shows the average PSD score compared with $k$. This is a flat line due to no influence from $k$. The PSD score remains at one for \textit{Naga}'s manufacturer assignment because all the nodes are always assigned from the same manufacturer. If any of the nodes are changed to a different manufacturer, more manufacturers will be in the path, leading to a penalty, and the average PSD score drops below one. However, if the flows are of different weights, the weights would influence the manufacturer assignment to prioritize the highest-weighted flows.

From Fig.~\ref{fig:special}, \textit{Naga} outperforms the centrality metrics-based manufacturer assignment even in these special cases.

%% file: 7_Conclusion.tex
\section{Conclusion}
\label{chap:conclusion}

Technology and data sovereignty created waves in the world of politics and engineering alike when several regions like Europe~\cite{sov1, sov3, sov4, sov5}, Middle East~\cite{middleeast}, Africa~\cite{africa}, and Asia~\cite{asia1, asia2} expressed concerns and motivation towards technology sovereignty. 
This work shows that network operators must consider network sovereignty along with other dependability attributes to build a robust network.
In this work, we laid the foundation for network sovereignty studies by investigating the challenges of establishing a sovereign network. We introduced a novel metric, the PSD score, to quantify network sovereignty. 
Our proposed metric supports the comparison of different manufacturer assignments and can lead to a better consideration of dependencies on manufacturers in network planning and operations. Moreover, we provide an ILP formulation called \textit{Naga} to maximize network sovereignty based on the PSD score. We evaluated \textit{Naga} and showed its superiority over centrality metric-based heuristics.

With more than three manufacturers, \textit{Naga} gives approximately $60\%$ increase in the PSD score compared with centrality metrics-based heuristics. However, the network operator is recommended to perform the manufacturer assignment based not only on PSD scores but also on network size, cost, component availability, manufacturer availability and trustworthiness, and other network-specific requirements. 

The immediate limitation of \textit{Naga} is its unsatisfying scalability in larger networks. Additionally, while using \textit{Naga} and the PSD score, the operator needs to choose an appropriate number of distinct paths. This key criterion can largely impact network sovereignty. The alternative is to consider a cut set-based approach for manufacturer assignment. This approach may mitigate both limitations. Moreover, from an operator's standpoint, our network sovereignty study combined with a network availability study will be complete and meaningful.

%% file: 0A_MAIN.bbl
\begin{thebibliography}{10}
\providecommand{\url}[1]{#1}
\csname url@samestyle\endcsname
\providecommand{\newblock}{\relax}
\providecommand{\bibinfo}[2]{#2}
\providecommand{\BIBentrySTDinterwordspacing}{\spaceskip=0pt\relax}
\providecommand{\BIBentryALTinterwordstretchfactor}{4}
\providecommand{\BIBentryALTinterwordspacing}{\spaceskip=\fontdimen2\font plus
\BIBentryALTinterwordstretchfactor\fontdimen3\font minus \fontdimen4\font\relax}
\providecommand{\BIBforeignlanguage}[2]{{%
\expandafter\ifx\csname l@#1\endcsname\relax
\typeout{** WARNING: IEEEtran.bst: No hyphenation pattern has been}%
\typeout{** loaded for the language `#1'. Using the pattern for}%
\typeout{** the default language instead.}%
\else
\language=\csname l@#1\endcsname
\fi
#2}}
\providecommand{\BIBdecl}{\relax}
\BIBdecl
\renewcommand{\BIBentryALTinterwordstretchfactor}{4}

\bibitem{samsung}
M.~J. Loveridge \emph{et~al.}, ``{Looking deeper into the Galaxy (Note 7)},'' \emph{Batteries}, vol.~4, no.~1, p.~3, 2018.

\bibitem{canada}
A.~Rudolph, ``{What is Log4j and Why Did the Government of Canada Turn Everything Off?}'' 2022.

\bibitem{equifax}
J.~Luszcz, ``Apache struts 2: how technical and development gaps caused the equifax breach,'' \emph{Netw. Secur.}, vol. 2018, no.~1, pp. 5--8, 2018.

\bibitem{Avizienis:2004}
A.~Avizienis \emph{et~al.}, ``Basic concepts and taxonomy of dependable and secure computing,'' \emph{{IEEE} Trans. Depend. Sec. Comput.}, vol.~1, no.~1, pp. 11--33, 2004.

\bibitem{edler2020technology}
J.~Edler \emph{et~al.}, ``{Technology sovereignty: From demand to concept [Technologiesouver{\"a}nit{\"a}t: Von der Forderung zum Konzept]},'' Fraunhofer Inst. for Syst. and Innov. Res. (ISI), Tech. Rep., 2020.

\bibitem{weber2018sovereignty}
A.~Weber \emph{et~al.}, ``Sovereignty in information technology,'' \emph{Secur., saf. and fair market access by openness and control of the supply chain. Karlsruhe: KIT-ITAS}, 2018.

\bibitem{bodin}
J.~Bodin \emph{et~al.}, \emph{Les six livres de la R{\'e}publique}.\hskip 1em plus 0.5em minus 0.4em\relax Chez Iacques du Puys Paris, 1986, vol.~6.

\bibitem{grant}
P.~Grant, ``Technological sovereignty: forgotten factor in the ‘hi-tech’razzamatazz,'' \emph{Prometheus}, vol.~1, no.~2, pp. 239--270, 1983.

\bibitem{review_data_sovereignty}
P.~Hummel \emph{et~al.}, ``Data sovereignty: A review,'' \emph{Big Data \& Soc.}, vol.~8, no.~1, p. 2053951720982012, 2021.

\bibitem{sov2}
J.~Pohle and T.~Thiel, ``Digital sovereignty,'' \emph{Practicing Sovereignty: Digit. Involvement in Times of Crises}, vol.~54, p.~47, 2021.

\bibitem{sov1}
V.~Reding, ``Digital sovereignty: {Europe} at a crossroads,'' \emph{Eur. investment bank Inst.}, 2016.

\bibitem{sov3}
L.~Floridi, ``The fight for digital sovereignty: What it is, and why it matters, especially for the {EU},'' \emph{Philosophy \& Technology}, vol.~33, no.~3, pp. 369--378, 2020.

\bibitem{sov4}
M.~Bauer and F.~Erixon, \emph{Europe's Quest for Technology Sovereignty: Opportunities and Pitfalls}.\hskip 1em plus 0.5em minus 0.4em\relax Eur. Centre for Int. Political Econ., 2020.

\bibitem{asia2}
R.~Creemers, ``China’s conception of cyber sovereignty,'' \emph{{Governing cyberspace: Behavior, power and diplomacy}}, pp. 107--145, 2020.

\bibitem{walter2021indigenous}
M.~Walter \emph{et~al.}, ``Indigenous data sovereignty in the era of big data and open data,'' \emph{Aust. J. of Social Issues}, vol.~56, no.~2, pp. 143--156, 2021.

\bibitem{carroll2019indigenous}
S.~R. Carroll \emph{et~al.}, ``{Indigenous data governance: strategies from United States native nations},'' \emph{Data Sci. J.}, vol.~18, 2019.

\bibitem{hill2014growth}
J.~Hill, ``{The growth of data localization post-Snowden: Analysis and recommendations for US policymakers and business leaders},'' in \emph{The Hague Inst. for Global Justice, Conf. on the Future of Cyber Governance}, 2014.

\bibitem{gao2021data}
\BIBentryALTinterwordspacing
H.~S. Gao, ``Data sovereignty and trade agreements: Three digital kingdoms,'' Oct. 2021. [Online]. Available: \url{https://ssrn.com/abstract=3940508}
\BIBentrySTDinterwordspacing

\bibitem{niceone_cyberspace}
Y.~Nugraha \emph{et~al.}, ``Towards data sovereignty in cyberspace,'' in \emph{2015 3rd Int. Conf. on Inf. and Commun. Technol. (ICoICT)}, 2015, pp. 465--471.

\bibitem{niceone_cyberspace2}
D.~Polatin-Reuben and J.~Wright, ``{An Internet with BRICS Characteristics: Data Sovereignty and the Balkanisation of the Internet.}'' in \emph{FOCI}, 2014.

\bibitem{chander2014data_nationalism}
A.~Chander and U.~P. L{\^e}, ``Data nationalism,'' \emph{Emory LJ}, vol.~64, p. 677, 2014.

\bibitem{litsurveysovereignty}
M.~Hellmeier and F.~von Scherenberg, ``A delimitation of data sovereignty from digital and technological sovereignty,'' \emph{ECIS Res. Papers}, 2023.

\bibitem{red}
\BIBentryALTinterwordspacing
R.~Bhardwaj, ``Network availability, redundancy, resilience, diversity: What’s the difference?'' [Online]. Available: \url{https://networkinterview.com/network-availability-redundancy-diversity/}
\BIBentrySTDinterwordspacing

\bibitem{div}
\BIBentryALTinterwordspacing
E.~Birman, ``How to avoid common network diversity disasters,'' 2020. [Online]. Available: \url{https://www.linkedin.com/pulse/how-avoid-common-network-diversity-disasters-eugene-birman/}
\BIBentrySTDinterwordspacing

\bibitem{redvsdiv}
\BIBentryALTinterwordspacing
{Advantage}, ``Network redundancy \& network diversity. what's the difference?'' 2019. [Online]. Available: \url{http://www.advantagecg.com/connection-advantage-blog/advantage-blog/2017/2/16/network-redunancy-and-diversity-whats-the-difference#:~:text=Redundancy%20%2D%20having%20two%20independent%20means,without%20sharing%20any%20common%20points.}
\BIBentrySTDinterwordspacing

\bibitem{astound}
\BIBentryALTinterwordspacing
{Grande Communications business}, ``The value of network diversity,'' \emph{Astound}. [Online]. Available: \url{https://mygrande.com/PDFs/The-Value-of-Network-Diversity-White-Paper-Grande.pdf}
\BIBentrySTDinterwordspacing

\bibitem{bauschert_vendor_selection}
R.~Romero-Reyes \emph{et~al.}, ``Impact of vendor selection on the total cost of ownership of intra-data centre networks,'' in \emph{21st Int. Conf. on Transparent Opt. Netw. (ICTON)}, 2019, pp. 1--4.

\bibitem{AugeOFC}
J.-L. Auge \emph{et~al.}, ``Open design for multi-vendor optical networks,'' in \emph{Opt. Fiber Commun. Conf. (OFC)}.\hskip 1em plus 0.5em minus 0.4em\relax Optica Publishing Group, 2019, p. Th1I.2.

\bibitem{gabilondo20215g}
A.~Gabilondo \emph{et~al.}, ``{5G SA multi-vendor network interoperability assessment},'' in \emph{IEEE Int. Symp. on Broadband Multimedia Syst. and Broadcasting (BMSB)}, 2021, pp. 1--6.

\bibitem{multi_vendor}
A.~Martinez \emph{et~al.}, ``Network management challenges and trends in multi-layer and multi-vendor settings for carrier-grade networks,'' \emph{{IEEE} Commun. Surveys Tuts.}, vol.~16, no.~4, pp. 2207--2230, 2014.

\bibitem{mine4}
S.~Janardhanan \emph{et~al.}, ``{Zohra: Joint Routing and Manufacturer Assignment Problem},'' in \emph{IEEE Int. Commun. Qual. and Rel. Workshop (CQR)}, Oct. 2023.

\bibitem{mine1}
S.~Janardhanan and C.~Mas-Machuca, ``Modeling and evaluation of a data center sovereignty,'' in \emph{18th Int. Conf. on the Des. of Rel. Commun. Netw. (DRCN)}, 2022, pp. 1--8.

\bibitem{mine2}
------, ``Modeling and evaluation of a data center sovereignty with software failures,'' in \emph{6th Int. Conf. on Syst. Rel. and Saf. (ICSRS)}, 2022, pp. 233--242.

\bibitem{china_website1}
P.~Leskin, ``{Here are all the major US tech companies blocked behind China’s ‘Great Firewall’},'' \emph{Business Insider}, vol.~10, 2019.

\bibitem{huawei_US_1}
\BIBentryALTinterwordspacing
{British Broadcasting Corporation}, ``\emph{US} telcos ordered to `rip and replace' {Huawei} components,'' \emph{BBC News}, Dec. 2020. [Online]. Available: \url{https://www.bbc.com/news/business-55269879}
\BIBentrySTDinterwordspacing

\bibitem{huawei_uk_1}
\BIBentryALTinterwordspacing
L.~Kelion, ``Huawei \emph{5G} kit must be removed from \emph{UK} by 2027,'' \emph{BBC News}, July 2020. [Online]. Available: \url{https://www.bbc.com/news/technology-53403793}
\BIBentrySTDinterwordspacing

\bibitem{spain_china_1}
\BIBentryALTinterwordspacing
J.~Masdeu, ``{El Gobierno lanza la convocatoria del 5G rural con cláusula antichina [The Government launches the call for rural 5G with anti-Chinese clause]},'' \emph{Telecommunicaciones, Economia}, Oct. 2023. [Online]. Available: \url{https://www.lavanguardia.com/economia/20231009/9285515/gobierno-lanza-convocatoria-5g-rural-clausula-antichina.html}
\BIBentrySTDinterwordspacing

\bibitem{uschina_chip1}
C.~P. Bown, ``{How the United States marched the semiconductor industry into its trade war with China},'' \emph{East Asian Econ. Rev.}, vol.~24, no.~4, pp. 349--388, 2020.

\bibitem{uschina_chip2_bloomberg}
\BIBentryALTinterwordspacing
D.~Wu, ``{TSMC Suspends Work for Chinese Chip Startup Amid US Curbs},'' \emph{Bloomberg Report}, Oct. 2022. [Online]. Available: \url{https://www.bloomberg.com/news/articles/2022-10-22/tsmc-said-to-suspend-work-for-chinese-chip-startup-amid-us-curbs}
\BIBentrySTDinterwordspacing

\bibitem{mine3}
S.~Janardhanan and C.~Mas-Machuca, ``Availability modeling and evaluation of switches and data centers,'' in \emph{Int. Conf. on Dependable Syst. and their Appl. (DSA)}, Aug. 2023.

\bibitem{main_all_values}
T.~A. Nguyen \emph{et~al.}, ``Reliability and availability evaluation for cloud data center networks using hierarchical models,'' \emph{{IEEE} Access}, vol.~7, pp. 9273--9313, 2019.

\bibitem{tsmc_clients_1}
R.~Abrams, ``{Asia Semiconductor Sector (Sector Review)},'' \emph{Asia Pacific Equity Res.}, Nov. 2013.

\bibitem{tsmc_clients_2}
B.~Wang, ``{Investment Recommendation- Taiwan Semiconductor Manufacturing Company, Ltd. (TSMC)},'' \emph{Investment Banking \& Asset Manage. - FINC356}, Dec. 2021.

\bibitem{vickybro1}
V.~Karunakaran \emph{et~al.}, ``{OpenROADM for Disaggregated Optical Networks: Challenges, Requirements and Evaluation},'' in \emph{Photon. Netw.; 24th ITG-Symp.}\hskip 1em plus 0.5em minus 0.4em\relax VDE, 2023, pp. 1--5.

\bibitem{Bur08}
W.~Burakowski \emph{et~al.}, ``Provision of end-to-end {QoS} in heterogeneous multi-domain networks,'' \emph{Annales des Télécommunications}, vol.~63, pp. 559--577, Dec. 2008.

\bibitem{Matos15}
F.~Matos \emph{et~al.}, ``Provisioning of inter-domain qos-aware services,'' \emph{J. of Comput. Sci. and Technol.}, vol.~30, pp. 404--420, Mar. 2015.

\bibitem{Durresi2010}
A.~Durresi \emph{et~al.}, ``Architecture for mobile heterogeneous multi domain networks,'' \emph{Mobile Inf. Syst.}, vol.~6, pp. 49--63, 2010.

\bibitem{Vilalta16}
R.~Vilalta \emph{et~al.}, ``{SDN/NFV} orchestration of multi-technology and multi-domain networks in cloud/fog architectures for {5G} services,'' in \emph{21st OptoElectronics and Commun. Conf. (OECC) held jointly with Int. Conf. on Photon. in Switching (PS)}, 2016, pp. 1--3.

\bibitem{Baranda18}
J.~Baranda \emph{et~al.}, ``Orchestration of end-to-end network services in the {5G}-crosshaul multi-domain multi-technology transport network,'' \emph{{IEEE} Commun. Mag.}, vol.~56, no.~7, pp. 184--191, 2018.

\bibitem{hansini1}
H.~Vijayaraghavan \emph{et~al.}, ``{Algorithmic and System Approaches for a Stable LiFi-RF HetNet Under Transient Channel Conditions},'' in \emph{IEEE 32nd Annu. Int. Symp. on Personal, Indoor and Mobile Radio Commun. (PIMRC)}, 2021, pp. 1048--1054.

\bibitem{Bernados16}
C.~Bernardos \emph{et~al.}, ``{5GEx}: Realising a {Europe-wide} multi-domain framework for software-defined infrastructures,'' \emph{Tran. on Emerging Telecommun. Technol.}, vol.~27, July 2016.

\bibitem{maria1}
M.~Samonaki \emph{et~al.}, ``Survivable node-disjoint routing in multi-domain networks,'' in \emph{IEEE Int. Conf. on Commun. (ICC)}, 2023.

\bibitem{Gao14}
C.~Gao \emph{et~al.}, ``Survivable inter-domain routing based on topology aggregation with intra-domain disjointness information in multi-domain optical networks,'' \emph{{IEEE} J. Opt. Commun. Netw.}, vol.~6, no.~7, pp. 619--628, Jul. 2014.

\bibitem{Gao11}
------, ``Domain-disjoint routing based on topology aggregation for survivable multi-domain optical networks,'' in \emph{IEEE Global Telecommun. Conf. (GLOBECOM)}, 2011, pp. 1--5.

\bibitem{Riti18}
R.~Gour \emph{et~al.}, ``Finding survivable routes in multi-domain optical networks with geographically correlated failures,'' \emph{{IEEE} J. Opt. Commun. Netw.}, vol.~10, no.~8, pp. 39--49, 2018.

\bibitem{gurobi}
\BIBentryALTinterwordspacing
{Gurobi Optimization, LLC}, ``{Gurobi Optimizer Reference Manual}.'' [Online]. Available: \url{https://www.gurobi.com}
\BIBentrySTDinterwordspacing

\bibitem{zoo}
S.~Knight \emph{et~al.}, ``The internet topology zoo,'' \emph{{IEEE} J. Sel. Areas Commun.}, vol.~29, no.~9, pp. 1765 --1775, Oct. 2011.

\bibitem{sndlib}
S.~Orlowski \emph{et~al.}, ``{SNDlib} 1.0—survivable network design library,'' \emph{Netw.: An Int. J.}, vol.~55, no.~3, pp. 276--286, 2010.

\bibitem{saxena2020centrality}
A.~Saxena and S.~Iyengar, ``Centrality measures in complex networks: A survey,'' \emph{arXiv preprint arXiv:2011.07190}, 2020.

\bibitem{sov5}
T.~A. Madiega, ``Digital sovereignty for {Europe},'' \emph{EPRS: Eur. Parliamentary Res. Service}, 2020.

\bibitem{middleeast}
\BIBentryALTinterwordspacing
M.~Soliman, ``{The Gulf has a {5G} conundrum and Open RAN is the key to its tech sovereignty},'' Jan. 2022. [Online]. Available: \url{https://www.mei.edu/publications/gulf-has-5g-conundrum-and-open-ran-key-its-tech-sovereignty}
\BIBentrySTDinterwordspacing

\bibitem{africa}
M.~Mawere and G.~van Stam, ``{Data Sovereignty: A Perspective From Zimbabwe},'' in \emph{12th ACM Conf. on Web Sci. Companion}, 2020, pp. 13--19.

\bibitem{asia1}
\BIBentryALTinterwordspacing
D.~Joshi, ``{Interrogating India's Quest for Data Sovereignty},'' July 2020. [Online]. Available: \url{https://ssrn.com/abstract=3648047}
\BIBentrySTDinterwordspacing

\end{thebibliography}
